\documentclass[useAMS,usenatbib]{mn2e}
\usepackage{graphicx}
\usepackage{subfigure}
\usepackage{amsmath}
\usepackage{amssymb}

\newcommand{\kel}{\mbox{ K}}
\newcommand{\mkel}{\mbox{ mK}}

\newcommand{\bq}{\begin{equation}}
\newcommand{\eq}{\end{equation}}
\newcommand{\bqa}{\begin{eqnarray}}
\newcommand{\eqa}{\end{eqnarray}}
\def\gsim{\;\rlap{\lower 2.5pt
 \hbox{$\sim$}}\raise 1.5pt\hbox{$>$}\;}
\def\lsim{\;\rlap{\lower 2.5pt
   \hbox{$\sim$}}\raise 1.5pt\hbox{$<$}\;}
\defcitealias{liu09a}{Liu et al. (2009a)}
\defcitealias{liu09b}{Liu et al (2009b)}

\title[Systematic Effects of Foreground Removal in 21cm Surveys of Reionization]{Systematic Effects of Foreground Removal in 21cm Surveys of Reionization} 
\author[N. Petrovic and S. P. Oh]{Nada Petrovic\thanks{E-mail: petrovic@physics.ucsb.edu} and S. Peng Oh\thanks{E-mail: peng@physics.ucsb.edu (corresponding author)} \\
Department of Physics; University of California; Santa Barbara, CA 93106, USA
}

\begin{document}
\bibliographystyle{mn2e}

\pagerange{000--000} \pubyear{0000}
\maketitle

\label{firstpage}

\begin{abstract}

21cm observations have the potential to revolutionize our understanding of the high-redshift universe. Whilst extremely bright radio continuum foregrounds exist at these frequencies, their spectral smoothness can be exploited to allow efficient foreground subtraction. It is well-known that--regardless of other instrumental effects--this removes power on scales comparable to the survey bandwidth. We investigate associated systematic biases. We show that removing line-of-sight fluctuations on large scales aliases into suppression of the 3D power spectrum across a broad range of scales. This bias can be dealt with by correctly marginalizing over small wavenumbers in the 1D power spectrum; however, the unbiased estimator will have unavoidably larger variance. We also show that Gaussian realizations of the power spectrum permit accurate and extremely rapid Monte-Carlo simulations for error analysis; repeated realizations of the fully non-Gaussian field are unnecessary. We perform Monte-Carlo maximum-likelihood simulations of foreground removal which yield unbiased, minimum variance estimates of the power spectrum in agreement with Fisher matrix estimates. Foreground removal also distorts the 21cm PDF, reducing the contrast between neutral and ionized regions, with potentially serious consequences for efforts to extract information from the PDF. We show that it is the subtraction of large-scales modes which is responsible for this distortion, and that it is less severe in the earlier stages of reionization. It can be reduced by using larger bandwidths. In the late stages of reionization, identification of the largest ionized regions (which consist of foreground emission only) provides calibration points which potentially allow recovery of large-scale modes. Finally, we also show that: (i) the broad frequency response of synchrotron and free-free emission will smear out any features in the electron momentum distribution and ensure spectrally smooth foregrounds; (ii) extragalactic radio recombination lines should be negligible foregrounds. 

\end{abstract}


\section{Introduction}
\label{section:intro}

21cm observations have the potential to revolutionize our understanding of the high-redshift universe (for recent reviews, see \citet{furl06-review,morales09}). By examining the imprints of the first galaxies on the intergalactic medium, they are complementary to (and potentially much more powerful than) direct surveys for proto-galaxies via Ly$\alpha$ and dropout techniques; furthermore, they can survey the universe on much larger scales. Moreover, unlike Ly-series absorption studies of high-redshift quasars, they constitute a fully three-dimensional probe which is not subject to saturation effects, and require no background sources. A host of upcoming low-frequency interferometers aim to detect fluctuations in 21cm emission from the Epoch of Reionization as a key science goal, including GMRT,\footnote{Giant
Metrewave Radio Telescope, http://www.gmrt.ncra.tifr.res.in/}
MWA,\footnote{Murchison Widefield Array,
http://www.haystack.mit.edu/ast/arrays/mwa/} LOFAR,\footnote{Low
Frequency Array, http://www.lofar.org/} 21CMA,\footnote{21 Centimeter
Array, http://web.phys.cmu.edu/\~{}past/} PAPER,\footnote{Precision
Array to Probe EoR, http://astro.berkeley.edu/\~{}dbacker/eor/}
SKA\footnote{Square Kilometre Array, http://www.skatelescope.org/}. 

One of the most difficult aspects of high-$z$ 21 cm measurements is the extreme brightness of the astronomical foregrounds relative to the cosmological signal (which has a brightness temperature $\delta T_{\rm b} \sim 20 \mkel$):  Galactic synchrotron emission has (at best) $T_{\rm gal} \sim 200$--$1000 \kel$ at these frequencies \citep{shaver99}  Extragalactic foregrounds, including radio galaxies, free-free emitters, and galaxy clusters \citep{dimatteo02, oh03, dimatteo04}, are also significantly brighter than the 21 cm signal.  Although these can be removed, the residual noise is large and numerous practical difficulties arise (see \S9.3 of \citealt{furl06-review}, and \S3.4 of \citealt{morales09}). Outstanding amongst these are terrestrial and satellite radio-frequency interference (RFI); ionospheric refraction and scintillation; and calibration errors in the direction-dependent gain and polarization response of the antennae. A particularly worrisome problem is the chromatic PSF of the instrument, which couples spatial fluctuations of spectrally smooth point sources into fluctuations in the frequency domain, an effect commonly dubbed "mode-mixing". Studies of the latter \citep{liu09a,bowman09,liu09b,datta10} show that while in principle this can be cleaned, the required calibration accuracy is formidable. By contrast, once the bright point sources are removed and the instrumental effects dealt with, removal of the spectrally smooth confused foreground of faint sources and Galactic emission is comparatively easy, and thought to be a solved problem \citep{zald04, morales04, santos05, wang06, morales05_foregrounds, mcquinn06}. In this paper, we focus {\it only} on this latter, comparatively 'easy' stage of continuum foreground removal. We show that there are systematic effects and biases--generally associated with the loss of large-scale power---which if not carefully taken into account, can lead to spurious results. 

The most widely discussed method for continuum foreground removal involves fitting and subtracting a smooth function to the data \citep{wang06,morales06,mcquinn06}. This ``trend removal" has a rigorous statistical justication as a means of extracting a tiny fluctuating signal from a huge, slowly varying background \citep{rybicki92}.  It corresponds to projecting out low-order, slowly-varying modes in the data \citep{mcquinn06}, with a price:  cleaning also attenuates the signal by removing its large-scale fluctuations. 
%This procedure assumes that the scales on which the signal and foregrounds have significant fluctuation power are sufficiently well-separated that the signal is unaffected by foreground removal (or, at least, that one is willing to sacrifice the smoothly varying modes in the signal).  
%%\citet{mcquinn06} have calculated the basic effects on the 21 cm power spectrum in the Fisher matrix approximation. %The order $n$ of a fitting polynomial plays an analogous role to the smoothing parameter in any regularization problem: if $n$ is too low, there are insufficient degrees of freedom to remove the foregrounds; if $n$ is too high, increasing amounts of cosmological signal are removed. 
It therefore severely restricts the spatial dynamic range of 21 cm observatories. For instance, MWA measurements of the 21cm power spectrum will be limited to roughly a decade in scale, from $k\sim 0.1-1 \, h\, {\rm Mpc}^{-1}$ \citep{lidz08}. Recent simulation studies have confirmed the efficacy of trend removal \citep{gleser08,jelic08,bowman-foreground-08,harker10}, and it is generally accepted as the best (and only mainstream) means of removing foregrounds. 

While the loss of large scale power is widely known and accepted as unavoidable, the systematic biases introduced by the process of foreground removal are much less well known. In this paper, we examine the consequences of removal of large-scale power in the line-of-sight direction in three key measures: (i) power spectrum. Foreground cleaning is often incorrectly characterized as removing all modes with $k < k_{\rm min}$. In fact, it removes all line of sight modes with $k_{\parallel} < k_{\rm min}$. We find that the converse of well-known aliasing effects \citep{kaiser91,lidz06} implies that removing large-scale power in radial modes removes small-scale power in transverse modes, creating an unavoidable bias across a range of scales (rather than just $k < k_{\rm min}$) in measurements of the 3D power spectrum. While this effect has been noticed before in simulations \citep{bowman09,harker09}, we provide a quantitative explanation and calculation of this effect. We show that this bias can be accounted for by correctly marginalizing over the large-scale modes along the line of sight, but at the expense of increased sample variance. We furthermore show that purely Gaussian realizations of the recovered power spectrum provide an accurate and extremely rapid means of generating Monte-Carlo error estimates for the power spectrum: fully non-Gaussian realizations are not necessary. (ii) PDF. The effect of foreground cleaning on the PDF has not been explored, apart from its effect on the skewness \citep{harker09}. We find that foreground cleaning causes unacceptable distortion of the one-point function. The PDF has surprisingly rich information about the progress and topology of reionization (\citet{furl04-21cmtop,wyithe07,harker09,ichikawa10}), but these will be obscured unless these artifacts can be removed. (iii) Tomography. Even the first generation of instruments will be able to image the very largest HII regions ($\gsim 20-30$ Mpc comoving), perhaps sourced by high-redshift quasars similar to those found by SDSS at $z\sim 6$. In principle, this might allow us to measure the neutral fraction by simply measuring the temperature contrast $\delta T_{b}$ between the HII region and the rest of the survey area (which contains many unresolved HII regions). We find that while the relative {\it shapes} of HII regions are preserved by foreground cleaning, the {\it temperature contrast} is significantly distorted, bedeviling efforts to measure $\delta T_{b}$ (see also \citet{geil08}). 

The outline of this paper is as follows. In \S\ref{section:method}, we describe our methodology for generating 21cm boxes, telescope noise, foregrounds, and removing foregrounds. In \S\ref{section:PS},\ref{section:tomography},\ref{section:PDF}, we discuss the effects of foreground removal on the power spectrum, tomography, and PDF respectively. In \S\ref{section:improvements}, we discuss possible improvements to foreground removal methods, and summarize conclusions in \S\ref{section:conclusions}. In Appendix \ref{section:smoothness}, we discuss bounds on the smoothness of the synchrotron and free-free continuum foregrounds. In Appendix \ref{section:RRL}, we discuss the integrated extragalactic radio recombination line background, and show that it is unlikely to be a show-stopper.  

\section{Method}
\label{section:method}

We now describe our simulations of 21cm signal, telescope noise, foreground, and the foreground removal process. It is important to note that this paper is squarely aimed at studying the biases introduced by subtraction of large-scale power which is ubiquitous to all foreground cleaning techniques. We do {\it not} aim at any advances in the state-of-the-art with regards to simulation or the cleaning process (except where it pertains to loss of large-scale modes; \S\ref{section:improvements}). Thus, our simulations and cleaning algorithm are relatively simple. We do not consider 'mode-mixing' and the consequences of bright source subtraction (\citetalias{liu09a,liu09b}; \citet{bowman09,datta10}); our foreground model is much less sophisticated than others (e.g., \citet{jelic08}); we largely ignore instrumental effects, such as leakage of polarized foregrounds to the total signal \citep{jelic10,geil10}; and we do not examine the subtleties of various foreground cleaning algorithms (e.g., \citep{harker09}). Indeed, often to make clear the biases that cleaning introduces, we omit noise from the plot, since many cleaning algorithms are linear processes (see discussion in \S\ref{section:fg_removal}). At this point, there are many detailed, 'end-to-end' simulation pipelines which incorporate all of these effects, and we have nothing to add to that literature. Rather, we aim at a theoretical understanding of the impact of the loss of large-scale power on various statistics, which can be understood robustly, independent of these further complications. Whilst many of these effects can be derived analytically, it is more transparent to show their effects on simulated maps.  

\subsection{The 21cm Signal}

To simulate 21cm signal, we use the semi-numeric simulation DexM \citep{mesinger07}, which can efficiently generate ionization maps of the IGM at high redshift far more rapidly (and for much larger boxes) than full numerical simulations, whilst showing excellent agreement with N-body simulations with full radiative transfer\footnote{Recently, an even more rapid variant of this algorithm, 21cmFAST \citep{mesinger10}, has been introduced, which achieves dramatic speed-ups by bypassing the halo finder.}. To briefly summarize their algorithm: 
3D realizations of linear density and velocity fields are first generated. Then, an excursion-set approach (\citet{bond91}, \citet{lacey93}) is used to filter the halos. The locations of the halos are adjusted using their first-order displacements obtained from the linear velocity field.  Finally, a similar filtering procedure is used to obtain an ionization field.  Unlike the halos, the bubbles are allowed to overlap and the excursion set barrier depends on ionization efficiency. The 21cm brightness temperature is then computed as:
\begin{eqnarray}
T_b(\nu)&=&\frac{T_{S}-T_{\gamma}}{1+z} (1-e^{-\tau_{\nu_{0}}}) \nonumber \\
	&\approx& 9(1+z)^{1/2} x_{\rm{HI}} (1+\delta_{nl}) \frac{H}{dv_r/dr+H} \, \rm{mK}
\end{eqnarray}     
where $T_S$ is the gas spin temperature, $\tau_{\nu_{0}}$ is the optical depth at the 21-cm frequency $\nu_{0}$, $\delta_{nl}$ is the physical overdensity, H is the Hubble parameter, $dv_{r}/dr$ is the comoving gradient of the line of sight component of the comoving velocity, and all the quantities are evaluated at redshift z=$\nu_{0}/ \nu-1$, under the approximations that $T_S>>T_{\gamma}$ and $dv_r/dr \ll H$ (i.e., redshift space distortions are ignored). Our fiducial $(200)^{3}$ pixel box--kindly provided to us by the authors---is 100 Mpc on a side (resulting in a cell size of 0.5 Mpc) at $z=8$ and a neutral fraction $x_{\rm HI}=0.56$. The bubble size distribution can be seen in Fig 7. of \citet{mesinger07} and is peaked at $\sim 8$ Mpc, with a significant tail out to $\sim 30$ Mpc.

We also generate 3D realizations of this box which have the same power spectrum but are Gaussian random fields.  These realizations do not contain ionized bubbles and have a different probability distribution function than the 21cm box; in particular there is no delta function distribution of fully ionized pixels. We found that the foreground subtraction affected the spherically averaged power spectrum and the average line of sight power spectrum similarly for both the initial 21cm box and the gaussian realization boxes.  Hence, we were able to use the realization boxes for Monte-Carlo calculations where many realizations with accurate power spectra are required, rather than repeating the entire semi-numeric process described above many times. More details of this are in \S\ref{section:gauss_box}. 

\subsection{Telescope Noise}
\label{section:noise}

As explained in \S\ref{section:fg_removal}, in the most widely used variants such as polynomial fitting, foreground cleaning is a linear operation. Thus, the impact of foreground cleaning on signal, foregrounds and noise can be considered separately. For the sake of clarity, we often do not simulate telescope noise, in order to clearly see the systematic biases foreground cleaning creates in the signal. Below, we describe how the telescope noise is simulated when we do consider it, in \S\ref{section:error}. 

For the sake of definiteness, we consider parameters appropriate to MWA, as specified in \citet{bowman09}. The MWA consists of $N_{\rm a}=500$ tiles, each with effective area $A_{\rm e} \approx {\rm min}(16,16 (\lambda^{2}/4 {\rm m^{2}}))$. We assume a smooth antenna density profile $n_{\rm a}(r) \propto r^{-2}$ within a 750m radius, with a core density of one tile per $36 \, {\rm m^{2}}$. We assume the sky-dominated system temperature to be $T_{\rm sys} \sim 440 [(1+z)/9]^{2.6}$K, and that a single field of view is observed for $t_{o} = 1000$hr (this is somewhat optimistic, but does not affect our basic conclusions). The baseline number density is then: 
\begin{equation}
n_{b}(U,\nu)=C_{b} \int_{0}^{r_{\rm max}} 2 \pi r {\rm d}r n_{\rm a}(r) \int_{0}^{2 \pi} {\rm d}\phi \, n_{\rm a}(|{\bf r}-\lambda{\bf U}|),
\end{equation}
where $C_{b}$ is a frequency dependent normalization (since $U \propto \nu$) such that $\int dU n_{b}(U,\nu)=N_{a}(N_{a}-1)/2$. The rms noise per visibility per frequency channel is then: 
\begin{equation}
\Delta V_{\rm N}({\bf U},\nu)=  \frac{\lambda^{2} T_{\rm sys}}{A_{e} \sqrt{\Delta \nu t_{\rm {\bf U}}}} \, {\rm K}. 
\label{eqn:noise}
\end{equation}
where $\delta \nu$ is the frequency bin width, and $t_{\rm {\bf U}}$ is the integration time for that visibility. The average integration time that an array observes the visibility ${\rm {\bf U}}$ is \citep{mcquinn06}:  
\begin{equation}
t_{\rm {\bf U}} \approx \frac{A_{e}}{\lambda^{2}} t_{o} n_{b}(U,\nu). 
\label{eqn:t_U}
\end{equation}
For each frequency channel, we draw complex visibilities from Gaussian distributions with rms given by equation (\ref{eqn:noise}), and respecting $V(U,\nu)=V^{*}(-U,\nu)$. We then inverse Fourier transform each slice to obtain the noise cube.  

\subsection{Foregrounds}
\label{section:foregrounds} 

Foreground brightness temperature fluctuations will be larger than the
cosmological 21-cm signal by several orders of magnitude. The three main
sources of foreground contamination of the 21-cm signal are Galactic
synchrotron (which comprises $\sim$\,70$\%$), extragalactic point
sources (27$\%$) and Galactic bremsstrahlung (1$\%$)
\citep{shaver99}. The frequency dependence of these foregrounds can
be approximated by power laws with a running spectral
index \citep{shaver99,tegmark00}. While the sum of power laws is
not in general a power law, over a relatively narrow frequency range
(such as that considered in this paper where $\Delta\nu/\nu\ll 1$), a
Taylor expansion around a power law can be used to describe the spectral shape. We therefore also
approximate the sum of foregrounds as a power law with a running
spectral index, and specialize to the case of Galactic synchrotron
emission, which dominates the foregrounds\footnote{The primary effect
of other foregrounds will be to slightly alter the spectral index and
the frequency dependence of the spectral index, and also (in the case
of unresolved radio point sources) the angular structure of the
foreground at small scales. Neither of these effects is important as
our foreground removal technique is not sensitive to the specific
value of the spectral index we adopt. As for angular fluctuations,
these correspond to zero-point fluctuations along different lines of
sight, which are immediately removed by foreground cleaning. We have
experimented with increasing angular fluctuations by a factor of 10,
and found little difference.}. For more detailed foreground models
including synchrotron emission from discrete sources such as supernova
remnants, and free-free emission from diffuse ionised gas, see
\citet{jelic08}. We assume that the brightest point sources have been
removed and only consider unpolarised foregrounds.

The intensity of the Galactic synchrotron emission varies as a
function of both sky position and frequency. We model the frequency and angular dependence of galactic
synchrotron foreground emission as follows. We first construct a realisation of the angular fluctuations in the foreground (around the mean brightness temperature) at a particular frequency $\nu_0$ using the relation
\begin{equation}
\frac{l^2 C_l(\nu_0)}{2\pi}=\left(\frac{l}{l_0}\right)^{2-\beta} T_{l_0}^{\rm syn}(\nu_0)^{2},
\end{equation}
where the frequency dependence of the fluctuation amplitude is given by
\begin{equation}
T_{l_{0}}^{\rm syn}(\nu_0)=A_{l_{0}}^{\rm syn} \bigg( \frac{\nu_0}{\rm 150\,MHz}\bigg)^{-\alpha_{\rm syn}-\Delta \alpha_{\rm syn} \rm log_{10}[\nu_0/(\rm 150\,MHz)]},
\end{equation}
where $\alpha_{\rm syn}, \Delta \alpha_{\rm syn}$ have mean values $\bar{\alpha}_{\rm syn}=2.55$, $\overline{\Delta \alpha}_{\rm syn}=0.1$, $\beta=2.5$ and $A_{l_{0}}^{\rm
syn}=25$\,K \citep{shaver99,tegmark00,wang06}. The latter two
values are extrapolated from 30 GHz CMB observations. To model spatial variations, we draw the parameters $\alpha_{\rm syn}$ and $\Delta \alpha_{\rm syn}$ from a gaussian distribution with standard deviation $\sim 10\%$ of the mean. 

We use the angular power spectrum to generate a two-dimensional
realisation of brightness temperature fluctuations $\Delta T^{\rm syn}(\mathbf{\theta})$ as a function of angular position $\mathbf{\theta}$ 
to which we add a mean sky brightness $\bar{T}^{\rm syn}=165(\nu_0/\mbox{185\,MHz})^{-2.6}$\,K (although
interferometers will generally not be sensitive to the temperature
zero-point). We then extend
this foreground plane into three dimensions by extrapolating along each line of sight:
\begin{equation}
T^{\rm syn}(\mathbf{\theta},\nu) = \left[\bar{T}^{\rm syn} + \Delta T^{\rm syn}(\mathbf{\theta})\right] \bigg( \frac{\nu}{\nu_0}\bigg)^{-\alpha_{\rm syn}-\Delta \alpha_{\rm syn} \rm log_{10}[\nu/\nu_0]},
\end{equation}
Our results are not sensitive to the details of the assumed foreground model. For instance, an increase in the amplitude of foreground brightness temperature fluctuations, by an order of magnitude, results in no change to our results after foreground cleaning. %Most of the error in foreground fitting comes from``noise" due to the instrument and the cosmological 21-cm signal.
The foreground completely swamps the HII region
signal, which is not visible in the combined maps. 
The mean and standard deviation of the foreground brightness (at the central frequency 158 MHz of these cubes) are $2.5\times10^5$\,mK and $1.4\times10^4$\,mK respectively, which should be compared to the  $\sim10$\,mK 21-cm fluctuations. Note that our simulations exclude the mean sky brightness, which cannot be measured by an interferometer.

%To obtain our final data cube, we perform a 2D Fourier transform on the combined signal plus foregrounds at each frequency, and delete all Fourier components where $n_{b}(U,\nu)=0$ (i.e., there are no baselines corresponding to these visibilities).  We then inverse Fourier transform each frequency slice, and add this to the noise cube, to obtain our final data cube. 

\subsection{Foreground Removal}
\label{section:fg_removal}

We employ here standard foreground cleaning techniques which have been promulgated in the literature \citep{wang06,morales06,mcquinn06,gleser08,jelic08,bowman09}: fitting a smooth function, such as a low-order polynomial, to the data, and subtracting it off, under the implicit assumption that the foreground along the line of sight varies much more slowly than the cosmological signal. There have been recent attempts to develop non-parametric techniques \citep{harker09}, which show promise, but at present yield no decisive advantage, whilst considerably complicating error analysis. We therefore only discuss standard parametric techniques. We also only consider foreground subtraction in image space, rather than Fourier space, since the two processes have been shown to yield fairly similar results (but note that such issues become more subtle once the frequency dependence of $uv$ coverage is taken into account (\citet{bowman09}; \citetalias{liu09a,liu09b}). 

Let us write the measured brightness temperature fluctuation in a pixel as:
\begin{equation}
x(\hat{\theta},\nu)=s(\hat{\theta},\nu) + n(\hat{\theta},\nu) + f(\hat{\theta},\nu)
\end{equation} 
where $s,n,f$ is the contribution from cosmological signal, telescope noise, and foregrounds respectively. In principle, our knowledge that the Galactic foreground is approximately a power-law might lead us to suppose that a fitting a log($\nu$)-log($x$) polynomial would be optimal, since that represents a Taylor expansion about a power law. In practice, we find that a $\nu-x$ polynomial gives very similar results. As in \citet{mcquinn06}, we therefore choose to work with these functions, since this implies that foreground cleaning is a {\it linear} process. %\citet{mcquinn06} work with orthogonal functions such as Legendre polynomials, where the algebra is particularly simple, but linearity is also true for non-orthogonal basis functions. 
We use a Gram-Schmidt procedure to form an orthogonal set of basis functions from the polynomial basis $1,x,...x^{n}$. 
Along each line of sight, our best fit to the foreground is: 
\begin{equation}
{\bf \bar{x}}={\bf P}{\bf q} 
\end{equation} 
where ${\bf \bar{x}}=\{\bar{x_{i}}\}$ is an n-dimensional vector consisting of the number of fitted frequency bins, ${\bf P}$ is an ${\it n \times m}$ matrix  composed of polynomials of order $\{0,1,...,m\}$ evaluated at frequencies $\{i \}$, and ${\mathbf q}$ is the $m$-dimensional vector of parameters (polynomial coefficients) to be estimated. If we now minimise $\chi^{2}=({\bf x}-{\bf \bar{x}})^{T}({\bf x}-{\bf \bar{x}})$, we obtain the simple linear estimator for the polynomial coefficients\footnote{Such a minimization implicitly assumes that the 'noise' to the fit--which is simply cosmological signal and telescope noise--is proportional to the identity matrix. This is approximately true for telescope noise--since different frequency slices are uncorrelated---but a poorer assumption for the cosmological signal. We defer further discussion of this issue to \S\ref{section:improvements}.}:
\begin{equation}
{\bf q}=({\bf P}^{T} {\bf P})^{-1} {\bf x}.
\end{equation}
The fit to the foreground is then:
\begin{equation}
{\bf \bar{x}}= {\bf P} ({\bf P}^{T} {\bf P})^{-1} {\bf x}.
\label{eqn:trend_removal}
\end{equation}
which implies that the foreground cleaned data is: 
\begin{equation}
{\bf \tilde{x}}={\bf x}-\bar{\bf x}=\left[1-{\bf P} ({\bf P}^{T} {\bf P})^{-1}\right] {\bf x} \equiv {\bf \Pi} {\bf x},
\label{eqn:cleaned_data}
\end{equation}
i.e. foreground cleaning can be represented by a linear projection operator ${\bf \Pi}$. In this case, $\tilde{\bf x}=\tilde{\bf s}+\tilde{\bf n}+\tilde{\bf f}$, and the covariance matrix of the cleaned data separates: 
\begin{equation}
\tilde{\bf C} = \langle {\bf x}^{T} {\bf x} \rangle = \tilde{\bf S} + \tilde{\bf N} + \tilde{\bf F}. 
\end{equation}
This is an extremely convenient property we will exploit, since it means we can consider the impact of foreground cleaning on the boxes of the cosmological signal alone, without worrying about interaction with the noise or foregrounds. We shall assume that the properties of the noise is fully known, so that $\tilde{\bf N}$ is fully specified a priori. 

For a linear foreground cleaning process such as equation (\ref{eqn:cleaned_data}), there are two possible sources of systematic biases. The first is that foreground cleaning is inefficient, and that the requirement that $\langle \tilde{\bf f} \tilde{\bf f}^{\dagger} \rangle \ll \langle \tilde{\bf s} \tilde{\bf s}^{\dagger} \rangle$ is not met; foregrounds residuals are still a significant contaminant. However, the expected spectrally smoothness of foregrounds (see Appendix \ref{section:smoothness} and \ref{section:RRL}) makes this unlikely. For the foregrounds we simulated, the cleaning procedure always left foregrounds residuals which were negligible compared to the signal. The second is that the process of cleaning itself, ${\bf s} \rightarrow \tilde{\bf s}$, introduces systematic biases in the signal. The latter is more serious, and can bias astrophysical and cosmological inferences, unless it is correctly understood.

Thus, we will generally illustrate the effects of foreground cleaning on the signal alone, without showing its impact on noise (note that foreground cleaning will have the same effect on noise as the signal) or foregrounds (which are always cleaned to a negligible level). We also do not depict the finite angular resolution of the telescope in our simulations. Whilst the effect of foreground cleaning remains the same (since it is a linear process), its impact is more difficult to assess visually in smoothed maps, although it clearly emerges in statistical measures such as the power spectrum. The effect of foreground cleaning and beam smoothing can be thought as of as a series of linear operators which alter the pristine signal; primarily on large scales and small scales respectively. Here, we focus on the former. 

\section{Effect of Foreground Removal on the Power Spectrum}
\label{section:PS}

\subsection{Aliasing}

\begin{figure}
%%%\epsscale{1.2}
\includegraphics[width=0.5\textwidth]{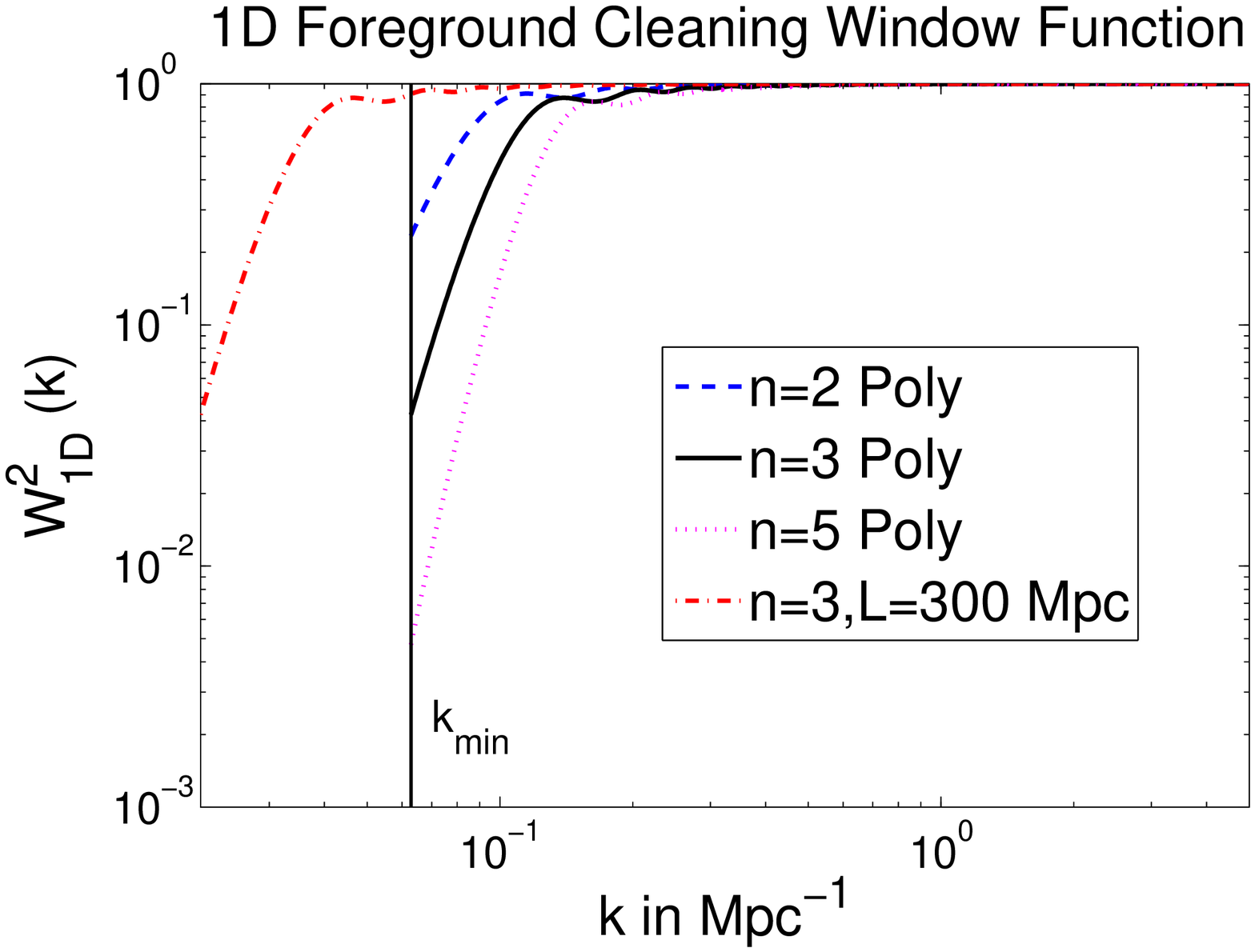} \\
\includegraphics[width=0.5\textwidth]{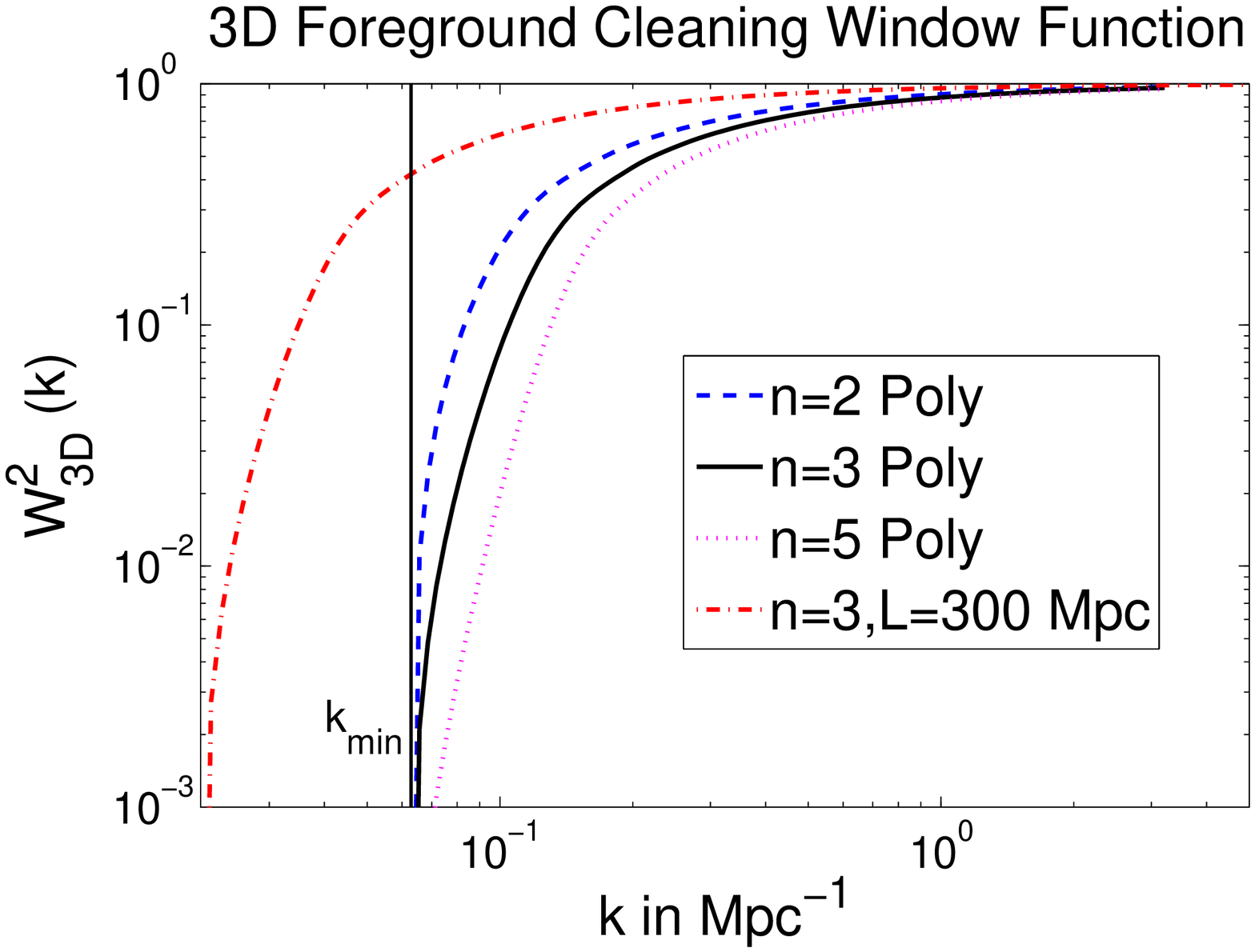}
\caption{Top panel: the 1D foreground cleaning window function, as given by equation (\ref{eqn:W_1D}). The wavenumber corresponding to the box size is marked by the line labeled $k_{\rm min}$. As the order of the polynomial increases, power at progressively smaller scales is subtracted from the box. Also shown is $W_{\rm 1D}^{2}$ for a box which has a bandwidth 3 times larger. It is identical to the window function for the smaller box, except that $k \rightarrow k/3$. Bottom panel: the 3D foreground cleaning window function, as given by equation \ref{eqn:W_3D}. Aliasing from the line of sight means that power is subtracted across a much wider range of scales. Otherwise, similar observations apply.}
\label{fig:window}
\end{figure}

Survey geometry and foreground cleaning modify the observed power spectrum. A Fourier mode with wavevector $\bf{k}$ will be convolved with a window function ${\bf W}(\bf{k},\bf{k^{\prime}}) = W_{\rm geom} W_{\rm fg-rm}$, where ${\bf W}_{\rm geom}$ is the window function due to the finite survey geometry, and ${\bf W}_{\rm fg-rm}$ encapsulates the effects of foreground removal. For a survey with cylindrical geometry $(L,R)$, where $L$ is the depth of the survey (given by the bandwidth $B$ of the survey) and radius $R$ is given by the field of view of the instrument, the survey window function is: 
\begin{equation}
W_{\rm geom} ({\bf q}) = \frac{2}{q_{r} R} j_{0}(q_{z} L/2) J_{1} (q_{r} R) 
\end{equation}
where ${\bf q}={\bf k}-{\bf k^{\prime}}$. This expression reflects the coupling between a wavevector ${\bf k}$ and all other wavevectors ${\bf k^{\prime}}$. In practice, survey geometry plays a much smaller role in modifying the power spectrum than foreground cleaning. 

What is the effect of foreground cleaning on Fourier modes? Let us assume that foreground cleaning is done on a voxel by voxel basis in the frequency direction, as in equation (\ref{eqn:cleaned_data}), i.e. its effect is equivalent to a linear projection operator ${\bf \Pi}$ operating in the $\hat{\bf z}$ direction. This cleaning process eliminates the orthonormality of Fourier modes $\mu_{k_{z}} \propto {\rm exp}(i k_{z} z)$. Following the notation of \citet{mcquinn06}, the new Fourier basis is $\tilde{\mu_{k}}= {\bf \Pi} \mu_{k_{z}}$, and the 1D window function is simply: 
\begin{equation}
W^{2}_{\rm 1D}(k_{z},k_{z}^{\prime})=\int_{0}^{L} {\rm d}z \tilde{\mu}_{k_{z}} \tilde{\mu}_{k_{z}^{\prime}}^{*}. 
\label{eqn:W_1D}
\end{equation}
We show this for foreground cleaning over an L=100 cMpc interval at z=8 in the top panel of Fig. (\ref{fig:window}), for polynomials of different order. Foreground cleaning causes a sharp reduction of power at scales comparable to the bandwidth over which the cleaning is done, $k_{\rm min} \approx 2 \pi/L$. As the order of the polynomial is increased, power on progressively smaller scales is removed. Also shown is $W_{\rm 1D}^{2}$ when foreground cleaning is performed over a bandwidth three times larger. It is identical to the window function for the narrower bandwidth, except that $k \rightarrow k/3$. Note that $W^{2}_{\rm 1D}(k_{z},k_{z}^{\prime})$ is actually a matrix; here we only depict the diagonal terms. In practice, the low sensitivity of first-generation experiments requires that the power spectrum can only be computed in a few large bins, and so the effect of mode-mode coupling is negligible.

\begin{figure}
%%%\epsscale{1.2}
\includegraphics[width=0.5\textwidth]{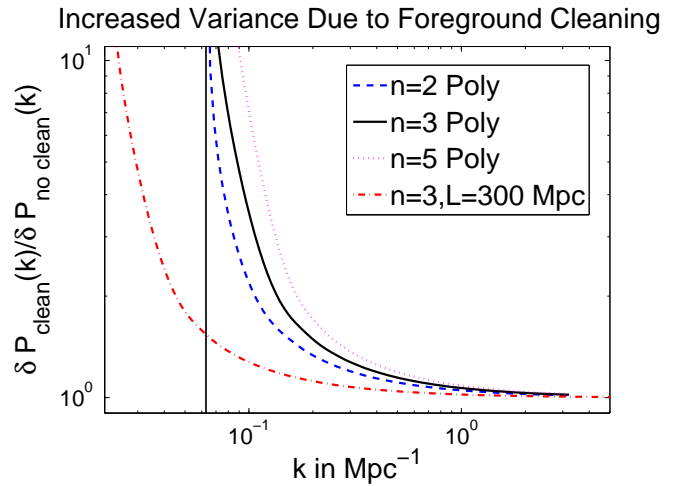}
\caption{The increase in the 1 $\sigma$ error on a power spectrum measurement, when foreground cleaning is performed, relative to an ideal measurement without foregrounds). For increasingly agressive foreground cleaning (i.e., a higher order polynomial), the number of measurable modes falls, and the variance of the measurement increases. The wavenumber corresponding to the box size is marked by the line labeled $k_{\rm min}$. Also shown is the increase in 1 $\sigma$ error for a box which has a bandwidth 3 times larger.}
\label{fig:increased_variance}
\end{figure}

How is the 3D angle-averaged power spectrum affected by foreground cleaning? It is worth reviewing some basics of power spectrum aliasing \citep{kaiser91}. The 1D power spectrum is simply a projection of the 3D power spectrum: 
\begin{equation}
P_{\rm 1D}(k_{\rm z}) \propto \int P_{\rm 3D}(k) dk_{x} \, dk_{y} \propto \int P_{\rm 3D} (\sqrt{k_{z}^{2} + k_{\perp}^{2}}) k_{\perp} dk_{\perp}  
\end{equation}
Thus, we obtain: 
\begin{equation}
\Delta^{2}_{\rm 1D}(k_{\rm z}) \propto k_{z} \int_{k_{\rm z}}^{\infty} \Delta^{2}_{\rm 3D}(k) k^{-2} dk
\end{equation}
where $\Delta^{2}_{\rm 1D}(k_{\rm z})= k_{z} P_{\rm 1D}(k_{\rm z})/\pi$, and $\Delta^{2}_{\rm 3D}(k)= k^{3} P_{\rm 3D}(k)/(2 \pi^{2})$. Thus, 1D modes receive contributes from {\it all} wavenumbers with $k \ge k_{\rm z}$. So small $k_{\rm z}$ modes can receive contributions from modes with large $k$ (which are oriented almost perpendicular to the line of sight). Conversely, power which is subtracted from small $k_{z}$ modes can affect the 3D power spectrum at large $k$.

\begin{figure}
%%%\epsscale{1.2}
\includegraphics[width=0.5\textwidth]{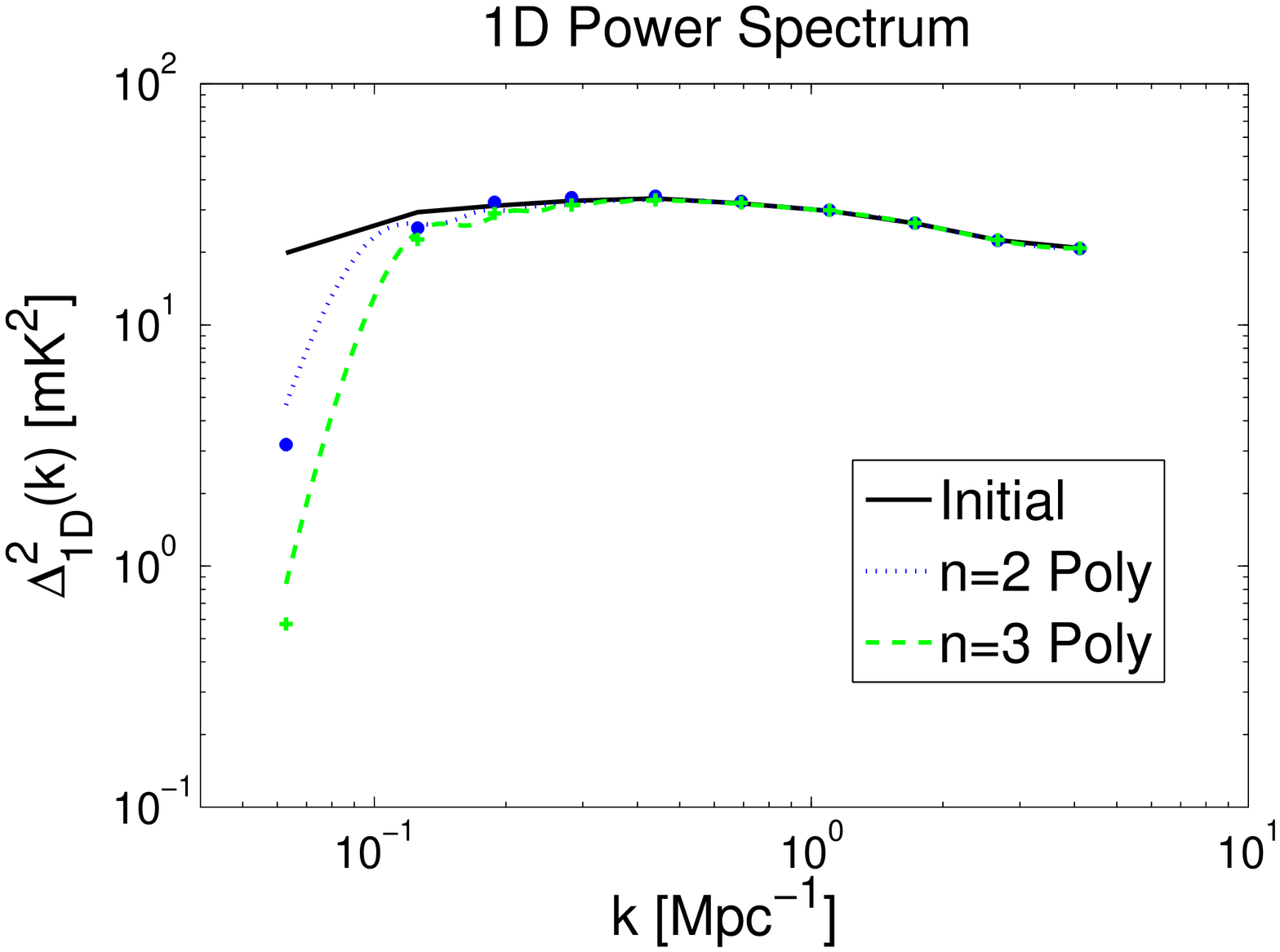} \\
\includegraphics[width=0.5\textwidth]{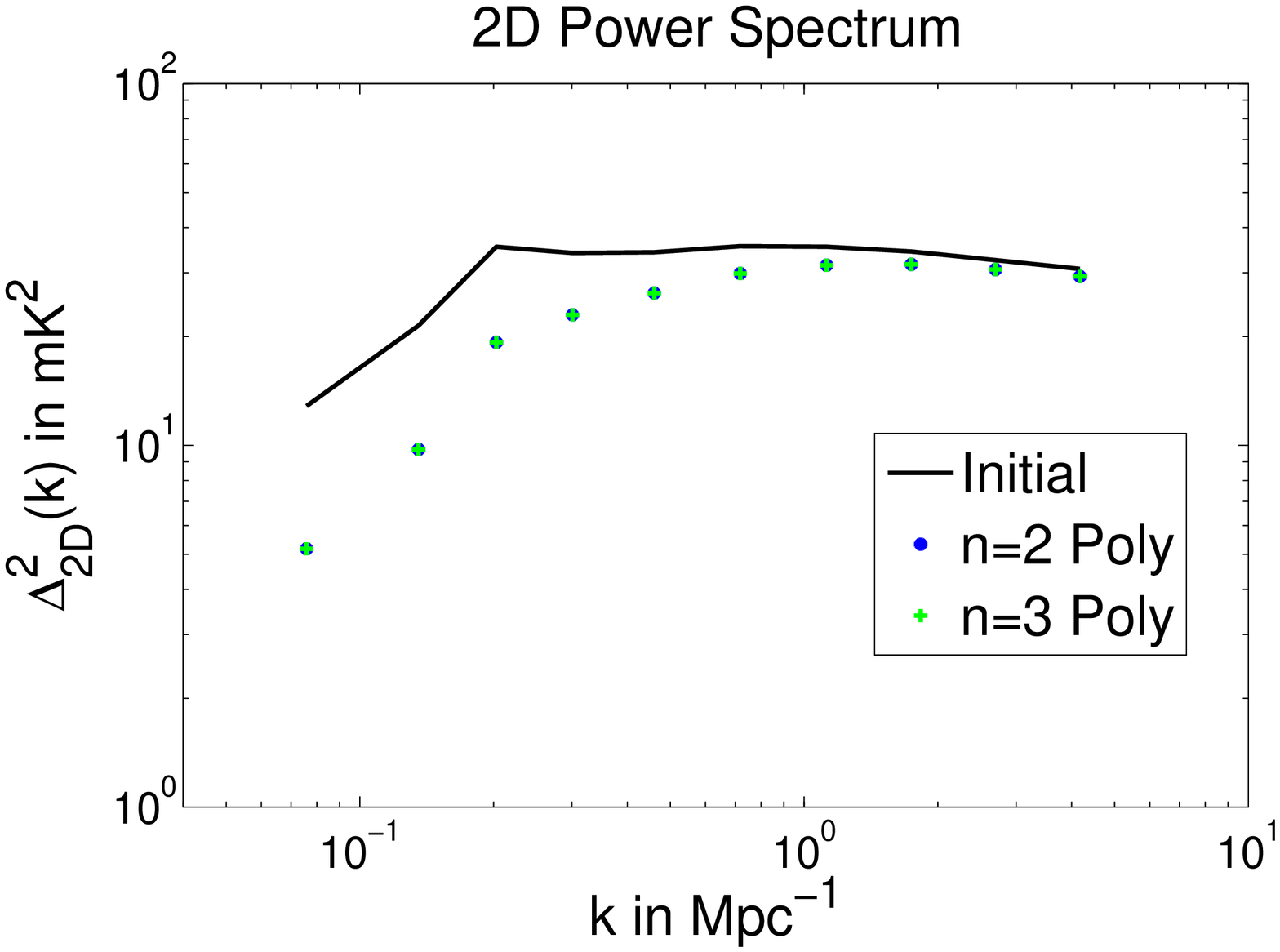} \\
\includegraphics[width=0.5\textwidth]{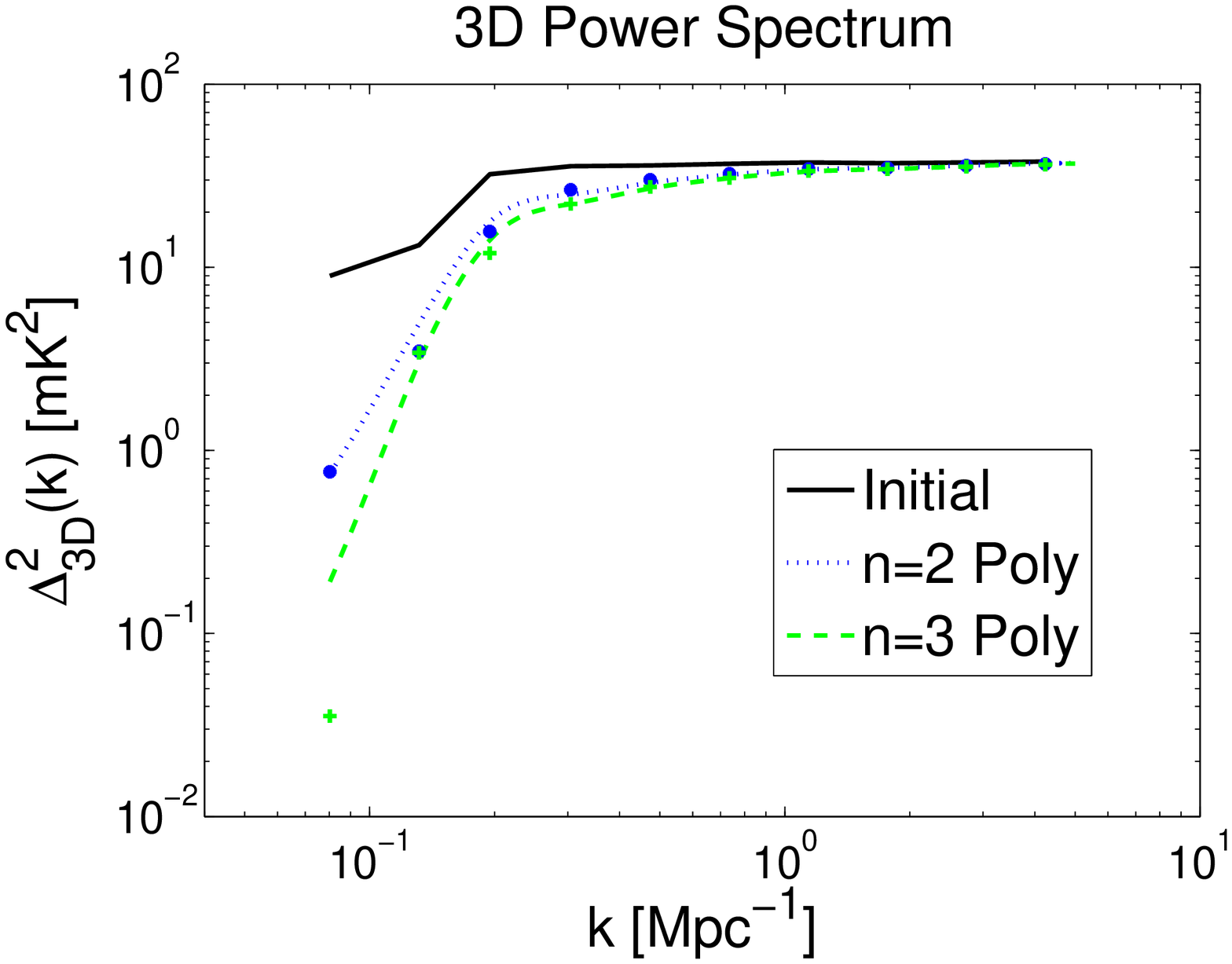}
\caption{The 1D, 2D and 3D power spectra after foreground cleaning by $n=2$, $n=3$ polynomials. The lines are the analytic calculation $\tilde{P}_{1D}(k) = W^{2}_{\rm 1D} P_{1D}(k)$, $\tilde{P}_{3D}(k) = W^{2}_{\rm 3D} P_{3D}(k)$, where $W^{2}_{\rm 1D}$, $W^{2}_{\rm 3D}$ is given by equation (\ref{eqn:W_1D}), (\ref{eqn:W_3D}) respectively. The points are the power spectra measured from the foreground subtracted box. The two show good agreement. Whilst foreground subtraction only affects large scales in 1D, aliasing results in broad suppression of power across a range of scales in 2D and 3D.}
\label{fig:PS}
\end{figure}

If we define $\mu= k_{z}/k$, then the window function for the 3D angle-averaged power spectrum is simply given by: 
\begin{equation}
W^{2}_{\rm 3D}(k) = \int W^{2}_{\rm 1D} (k \mu)  \, {\rm d}\mu.
\label{eqn:W_3D}
\end{equation}
This is shown in the bottom panel of Fig. (\ref{fig:window}), for our box of 100 cMpc (results corresponding to the flattened geometry of surveys will be shown in \S\ref{section:error}). The scaling properties are very similar to those of $W^{2}_{\rm 1D}$, except that power is subtracted across a much wider range of scales, as expected from the discussion above. 

Thus, foreground cleaning causes suppression of power across a wide range of scales in the angle-averaged 3D power spectrum. We will see how this can be dealt with in \S\ref{section:error}. However, even if the systematic bias can be appropriately dealt with, the elimination of low-wavelength modes in the radial direction unavoidably reduces the number of modes available to measure the power spectrum at a given wavenumber. The fraction of measurable modes is given by $W^{2}_{\rm 3D}(k)$; hence, sample variance is increased by foreground cleaning by a factor $1/\sqrt{W^{2}_{\rm 3D}(k)}$. We show this in Fig. (\ref{fig:increased_variance}). As expected, for increasingly agressive foreground cleaning (i.e., a higher order polynomial), the number of measurable modes falls, and the variance of the measurement increases. Thus, in order to measure a given wavenumber $k$, one would have to perform foreground cleaning over a bandwidth corresponding to a significantly larger lengthscale than $L = 2 \pi/k$. This increase in sample variance due to foreground cleaning is also evident in the Fisher matrix formalism of \citet{mcquinn06}.

How accurate are these analytic expectations? In Fig \ref{fig:PS}, we compare the power spectra predicted by these formulae (lines), to the power spectra measured directly from the box after foreground cleaning (points). For completeness, we also show the impact of foreground cleaning on the 2D power spectrum in the plane of the sky. The analytic and measured power spectra show good agreement. In particular, whilst foreground subtraction only affects large scales in 1D, aliasing results in broad suppression of power across a range of scales in 2D and 3D. 

Note that in making these comparisons, no foreground has been previously added to the 21cm box. If foregrounds are initially added, the power spectrum for $n=3$ polynomial subtraction is unchanged, but for $n=2$ polynomial subtraction, the power spectrum blows up at large scales: foreground cleaning is insufficiently aggressive. As emphasized by \citet{mcquinn06}, one always chooses the lowest order polynomial that is able to remove foregrounds well below the cosmological signal. 

\begin{figure}
%%%\epsscale{1.2}
\includegraphics[width=0.5\textwidth]{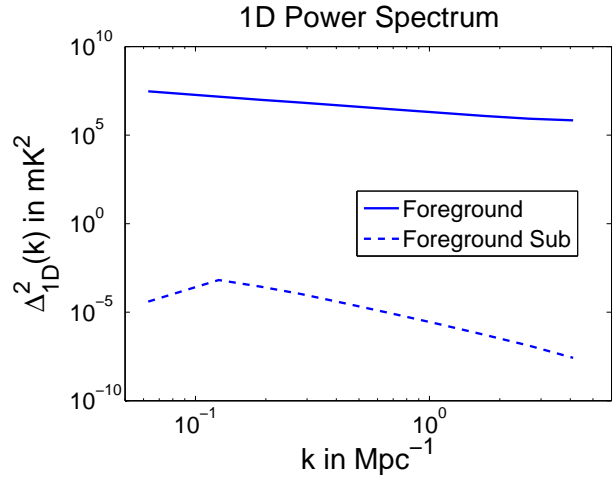} \\
\caption{The effect of foreground cleaning on the power spectrum of the foreground itself. Unlike for the signal (and noise) power spectra, fitting a smooth function such as a polynomial suppresses power on all scales for the smooth foreground. These effects also carry on to 3D (not shown).}  
\label{fig:ps_fg_noise}
\end{figure}

Given these considerations, one might be tempted to foreground clean by applying a sharp step function in $k_{z}$ space, rather than fitting a polynomial. The sharp decay in power at low wavenumber due to foreground cleaning by a polynomial might more simply be accomplished by excluding all modes with $k_{\rm z} < k_{\rm z, cut}$, where $k_{\rm z, cut}$ is some critical wavenumber above which the foreground has little power. Indeed, we shall shortly see that marginalizing over modes with $k_{\rm z} < k_{\rm z, cut}$ can eliminate systematic bias in power spectrum estimation. However, we should keep two facts in mind. Firstly, the (close to power-law) foreground power spectrum cannot be fully represented by a few Fourier modes in the line of sight; there is still power at high $k_{z}$. Secondly, the effect of foreground cleaning can be represented by a window function $\tilde{P_{\rm clean}} (k)= W^{2}(k) P(k)$ only for a field which is reasonably close to random phase; applying a 'matched filter' such as a smooth polynomial to spectrally smooth foregrounds brings about a much greater reduction of power on all scales. These effects are shown in Fig \ref{fig:ps_fg_noise}. Note that foreground cleaning can be represented as $\tilde{P}_{\rm clean}(k)= W^{2}(k) P(k)$ for the noise, which is a Gaussian random-phase field. 

%Another possible way to foreground clean would be to fit a polynomial, or other smooth function, to Fourier coefficients frequency direction. The relative merits of cleaning in real or Fourier space has been explored by other authors (e.g., \citet{}; it appears there is a slight advantage to cleaning in Fourier space), and we shall not explore this further.  

\subsection{Insensitivity to Non-Gaussianity of Signal}
\label{section:gauss_box}

The best means of performing error analysis for the complex data analysis pipeline in 21cm experiments is to perform Monte-Carlo simulations; in this way both systematic and statistical errors introduced at various stages can be fully understood. It is time-consuming to produce a new simulation box of 21cm signal for each Monte-Carlo realization, as is necessary to take cosmic variance into account. While extremely fast means of creating approximate 21cm realizations for a given ionization parameter now exist \citep{mesinger10}, these cannot be tuned to accurately represent different realizations of some specified underlying power spectrum, rather than the power spectrum of the particular theoretical model used to generate the box. The easiest way to go about this would be to simply create Gaussian realizations of a given power spectrum. 

However, it is not clear whether foreground cleaning will have the same impact on the power spectrum of a Gaussian and non-Gaussian field (such as a 21cm box). Least squares is an optimal means of fitting a parametric function when the residuals are Gaussian; however, it can be sub-optimal when residuals are non-Gaussian (for instance, when there are outlier data points which are highly unlikely in a Gaussian distribution; these will disproportionately affect the fit). In this case, the residuals, which constitute the 21cm signal itself, are non-Gaussian. In particular, there is strong phase coherence across ionized regions, and sharp features due to the boundary between ionized and neutral regions. 

\begin{figure}
\includegraphics[width=0.5\textwidth]{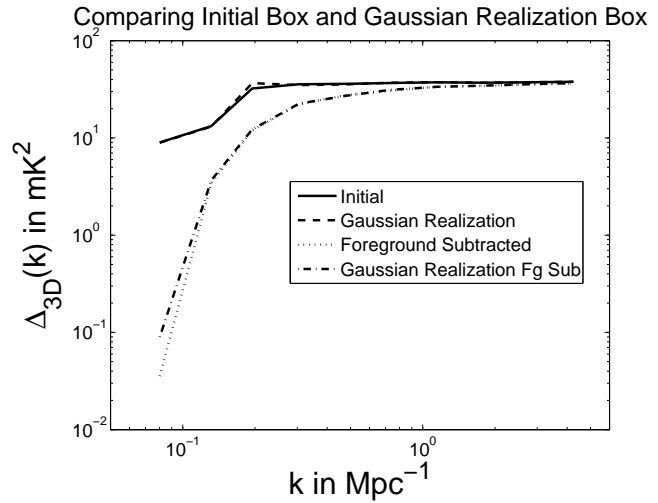}
\label{fig:gauss}
\caption{The effect of foreground subtraction on the power spectrum of the full 21cm box with non-Gaussian signal and on Gaussian realizations of this box, which have identical power spectra.  For the Gaussian case the power spectrum is averaged over 10 realizations, to reduce sample variance. Foreground subtraction has identical impact on the power spectra of the full 21cm box and Gaussian realizations, which implies that Gaussian realizations are adequate for rapid Monte-Carlo simulations.}
\label{fig:gauss_nongauss}
\end{figure}

In Fig \ref{fig:gauss_nongauss}, we compare the effect of foreground subtraction on the power spectrum of the full 21cm box with non-Gaussian signal and on Gaussian realizations of this box, which have identical power spectra.  For the Gaussian case the power spectrum is averaged over 10 realizations, to reduce sample variance. Foreground subtraction has identical impact on the power spectra of the full 21cm box and Gaussian realizations, which implies that Gaussian realizations can be used for rapid Monte-Carlo simulations. The reason for this identical impact is likely because non-Gaussianity is important on scales smaller than those affected by foreground cleaning. This is also the reason why the analytic calculation $\tilde{P}_{\rm clean} (k)= W^{2}_{\rm 3D} (k) P(k)$, where $W^{2}_{\rm 3D} (k)$ is given by equation (\ref{eqn:W_3D}), closely corresponded to the box simulations in Fig \ref{fig:PS}. 

\begin{figure}
%%%\epsscale{1.2}
\includegraphics[width=0.5\textwidth]{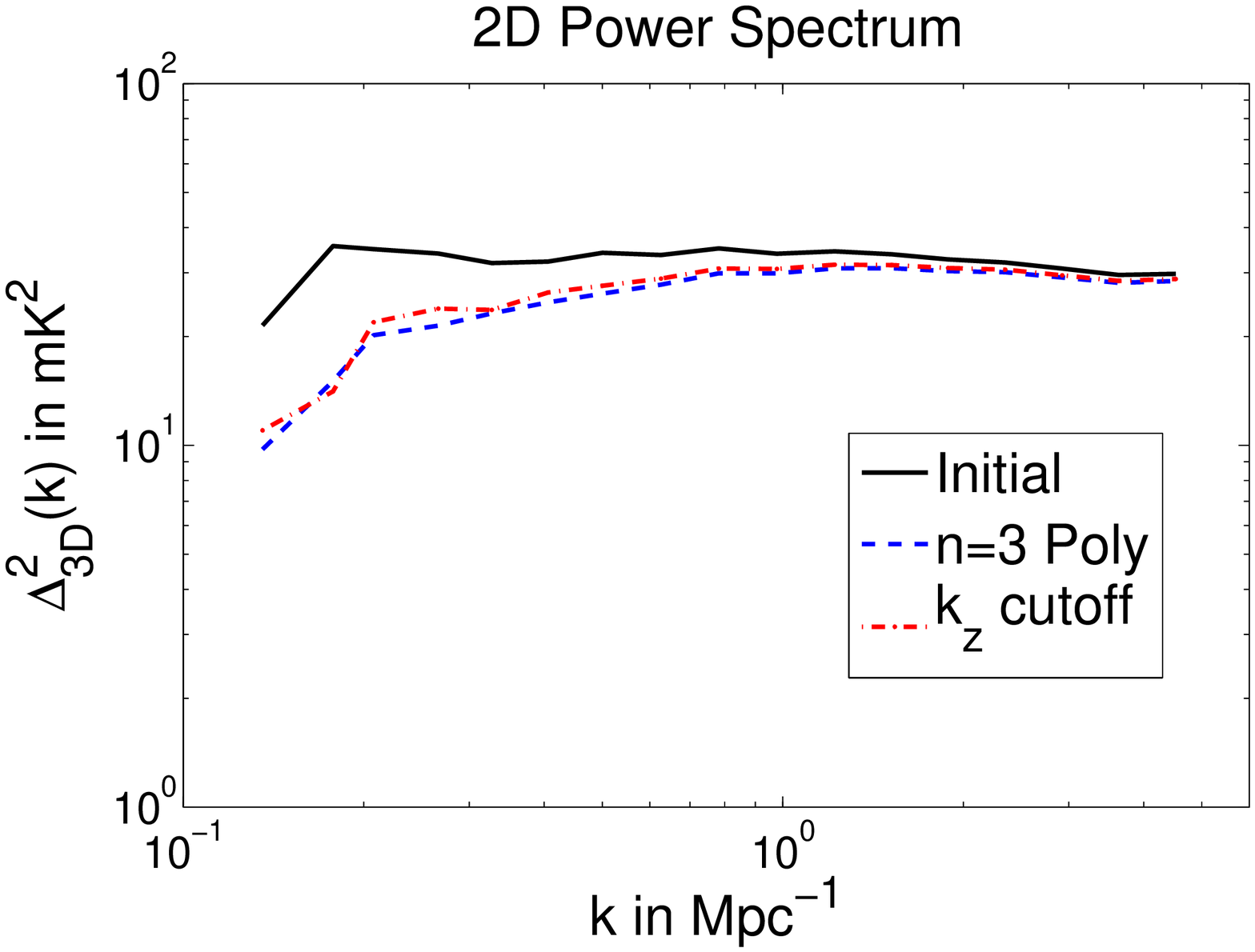} \\
\includegraphics[width=0.5\textwidth]{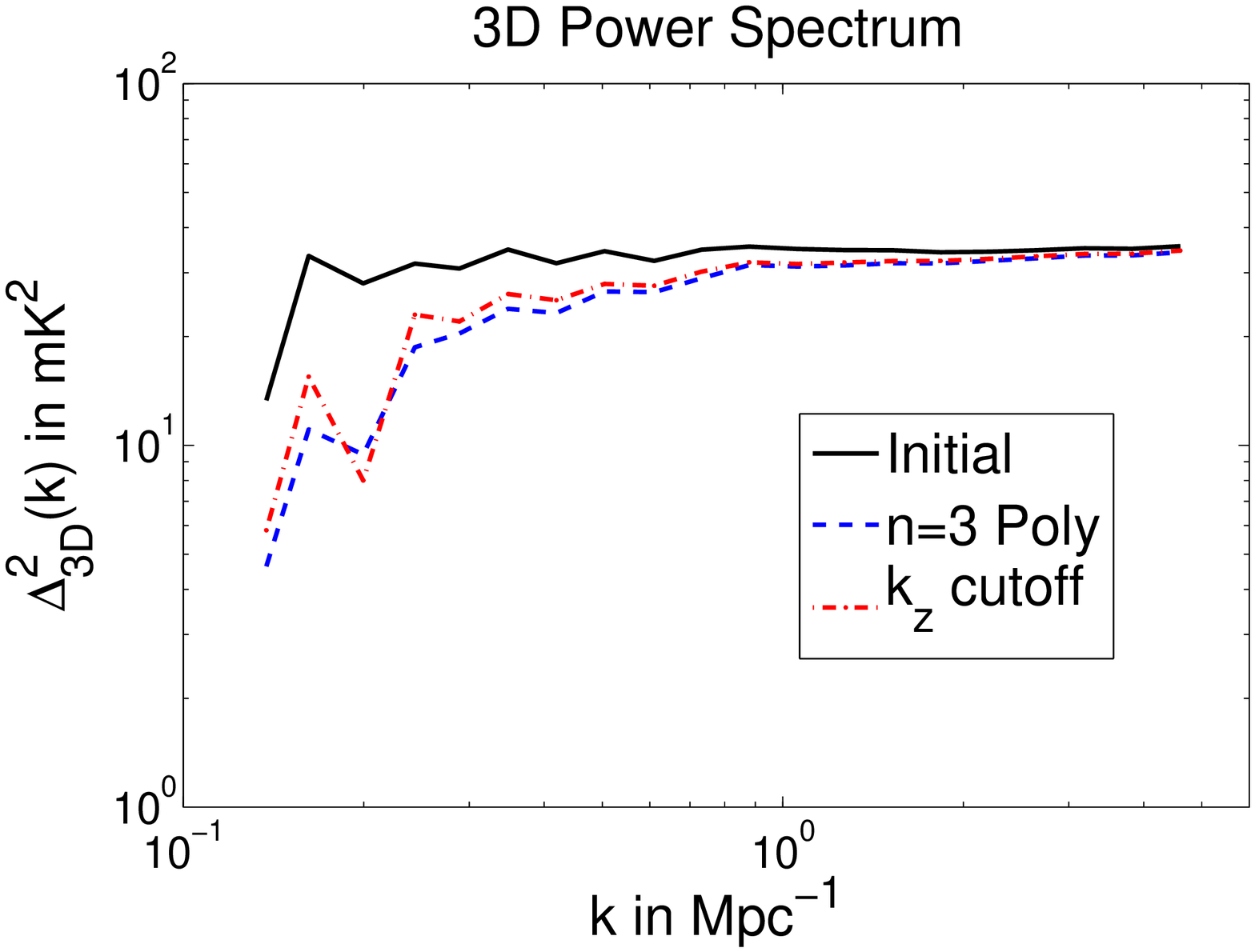} \\
\caption{The 2D and 3D power spectra after all modes with $k_{z} \leq 0.15 {\rm Mpc^{-1}}$ are set to zero; from Fig. \ref{fig:PS}, this is the wavenumber at which the 1D power spectra starts plummeting rapidly after foreground cleaning, for an $n=3$ polynomial. Indeed, such a sharp k-space cutoff mimics the broad suppression of power in 2D and 3D extremely well.}  
\label{fig:k_z_cutoff}
\end{figure}

\subsection{Eliminating Systematic Bias; Error Estimates}
\label{section:error}

\begin{figure}
%%%\epsscale{1.2}
\includegraphics[width=0.5\textwidth]{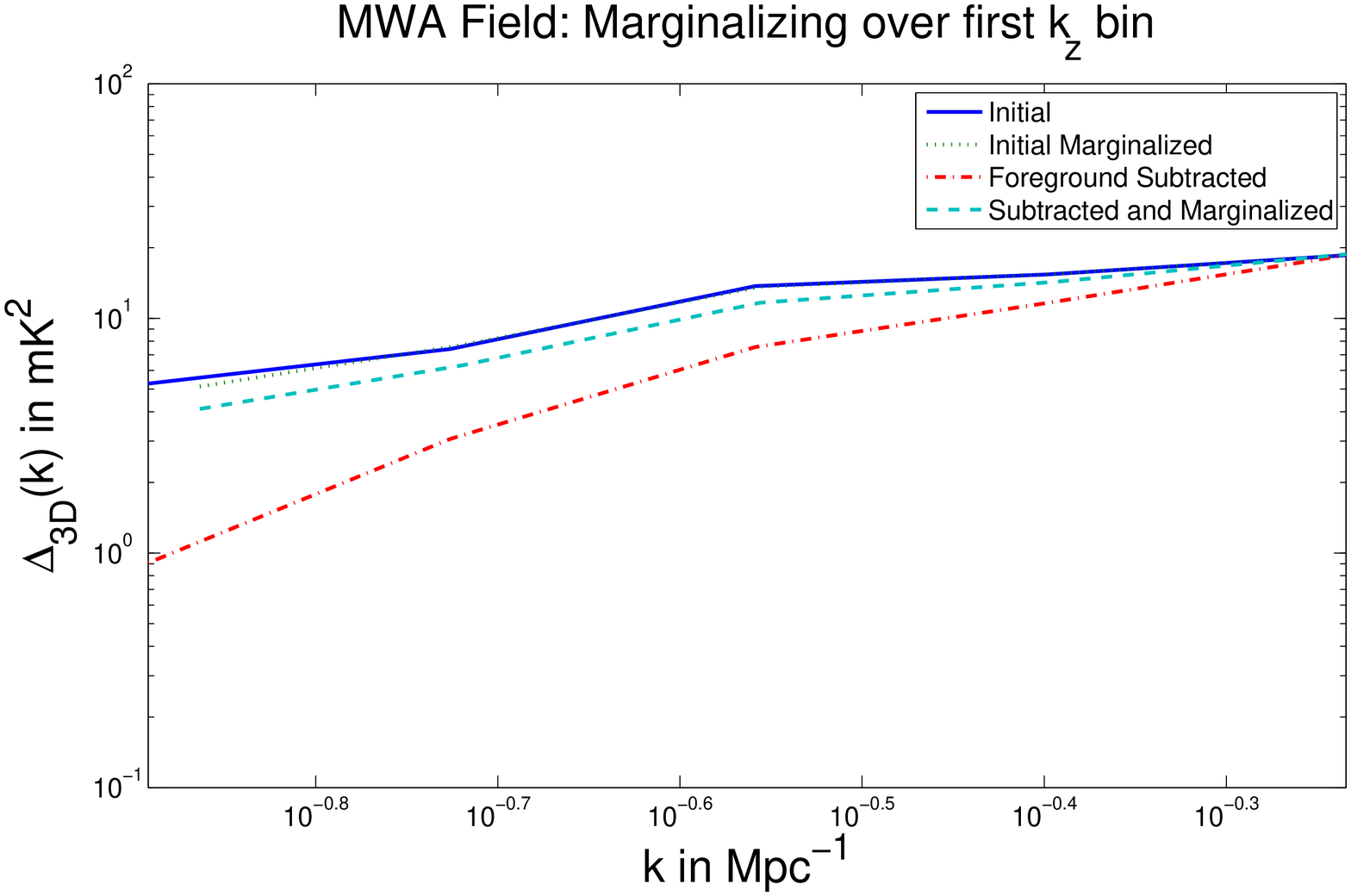} \\
\includegraphics[width=0.5\textwidth]{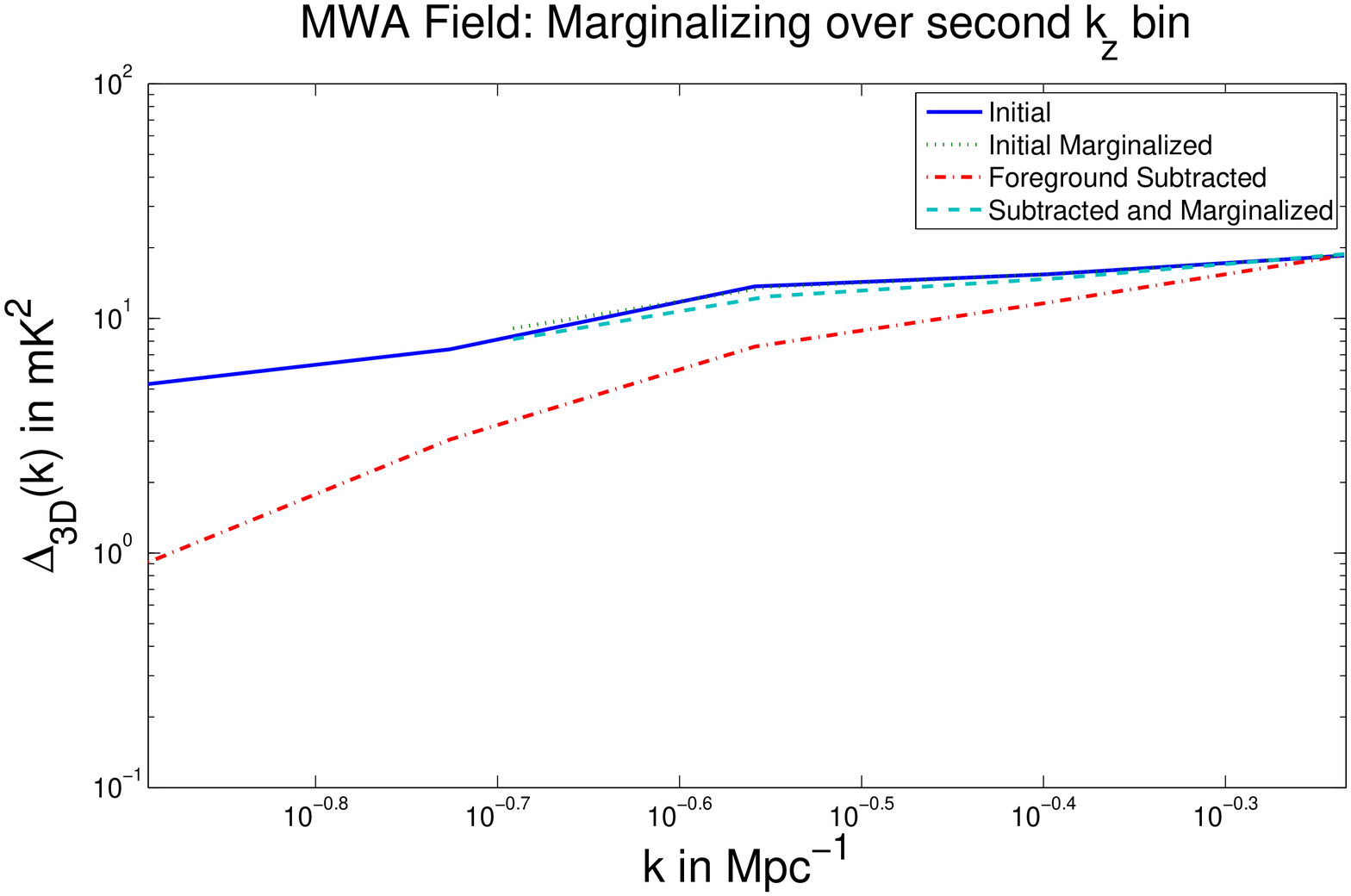}
\caption{Removing the systematic effects due to foreground cleaning by marginalizing over modes which have $k_{\rm z} < k_{\rm z,crit}$. The calculation is performed for survey geometry appropriate to the MWA, in 5 logarithmically spaced bins, and only modes with $k_{\rm z} > k_{\rm z,crit}$ are included in the calculation of the power spectrum. In the top panel, we marginalize over the first $k_{z}$ bin; much of the systematic bias is eliminated. In the lower panel, we marginalize over the first two $k_{z}$ bins. Although all of the systematic bias is removed, the power spectrum is measured over a smaller range of scales.}
\label{fig:MWA_marginalize}
\end{figure}

As previously discussed, excising modes with $k < k_{\rm z,crit}$ will have a similar effect on the signal and noise power spectra (though not on the foreground power spectrum) as foreground cleaning via fitting and subtracting a polynomial. In this case, $W_{\rm 1D}^{2}$ will simply be a step function: $W_{\rm 1D}^{2}=0$ for $k \leq k_{\rm z,crit}$, $W_{\rm 1D}^{2}=1$ for $k > k_{\rm z,crit}$. From equation (\ref{eqn:W_3D}), this implies that $W_{\rm 3D}^{2}(k) \approx 1 - \mu_{\rm crit} = (1-k_{\rm z,crit}/k)$. For a judiciously chosen $k_{\rm crit}$, this approximates the effect of polynomial fitting quite well. In Fig \ref{fig:k_z_cutoff}, we show the effect of mode excision for modes with $k_{z} \leq 0.15 \, {\rm Mpc^{-1}}$ on our 21cm box; from Fig \ref{fig:PS}, this is the wavenumber at which the 1D power spectra starts plummeting rapidly after foreground cleaning, for an $n=3$ polynomial. As can be clearly seen, a sharp k-space cutoff mimics the broad suppression of power in 2D and 3D extremely well. Thus, if after foreground cleaning we simply use uncontaminated modes with $k > k_{\rm z,crit}$ to estimate the power spectrum (i.e., we marginalize over modes with $k \lsim k_{\rm z,crit}$), we should have an {\it unbiased} estimator of the power spectrum, even at low $k$. At the same time, as seen in Fig. \ref{fig:increased_variance}, there will be unavoidably larger variance in the power spectrum estimate, since there are fewer independent modes sampling a given wavenumber. 

We test these ideas out on a simulated field of view appropriate to the MWA, which is much more extended in the plane of the sky than in redshift space. We consider an MWA field which is $\sim 800 \, {\rm deg^{2}}$ and $\sim 6$ MHz at $z=7.3$; this corresponds to a rectangular box 4350 cMpc by 4350 cMpc by 100 cMpc. To generate the simulated field, we use the fact that foreground cleaning is insensitive to the non-Gaussianity of the signal (\S \ref{section:gauss_box}), and simply generate a Gaussian random realization of a given power spectrum. We use the power spectrum of \citet{lidz08} at $z=7.32$, $x_{\rm HI}=0.46$, which is derived from radiative transfer simulations in a 130 Mpc $h^{-1}$ box. MWA will only be sensitive to power spectrum measurements over roughly a decade in scale, $k \sim 0.1 - 1 \, h \, {\rm Mpc^{-1}}$. Measurements even over this limited wavenumber range are valuable, since the redshift evolution of the amplitude and slope of the power spectrum contains valuable information \citep{lidz08}. The lower limit is given by foreground cleaning considerations.
%, since large scale power is removed by foreground cleaning. While this can be ameliorated by cleaning over larger bandwidths, over larger bandwidths, the effects of cosmic evolution then become important. 
The upper limit is given by telescope resolution: on small scales, the telescope samples relatively few modes (mostly along the line of sight) due to finite angular resolution, and the noise blows up. Given these considerations, we divide our box into 512 by 512 by 32 cells, which corresponds to cells of size 8.5 by 8.5 by 3.1 cMpc. Finer resolution is unnecessary over this range of wavenumber.

We divide the wavenumber range into 5 logarithmically spaced bins. In Fig. \ref{fig:k_z_cutoff}, we show how the systematic suppression of power due to foreground cleaning can be correctly accounted for by marginalizing over the appropriate modes, i.e., only including modes with $k_{\rm z} > k_{\rm z,crit}$ in the calculation of the power spectrum. In the top panel, we marginalize over the first $k_{z}$ bin; much of the systematic bias is eliminated. In the lower panel, we marginalize over the first two $k_{z}$ bins. Although all of the systematic bias is removed, the power spectrum is measured over a somewhat smaller range of scales. %Overall, this confirms that a simple marginalization procedure is able to handle systematic effects of foreground cleaning on the power spectrum. 

Let us now run more detailed calculations to confirm that such a marginalization procedure is able to return unbiased power spectrum estimates with the minimum variance error bars expected from a Fisher matrix calculation. For the latter, we use the expressions derived by \citet{mcquinn06}. For wavenumbers $k_{z},k_{z}^{\prime}$ at which the foregrounds can be cleaned well below the signal, the Fisher matrix is: 
\begin{equation}
F_{k_{z},k_{z}^{\prime}}^{{\bf k}_{\perp}} \approx w^{2} \left[ \frac{\tilde{\bf \mu}_{k_{z}}^{\dagger} \tilde{\bf \mu}_{k_{z}^{\prime}}}{T_{N}^{2}} - \frac{(\tilde{\bf \mu}_{k}^{\dagger} {\bf f})({\bf f}^{\dagger}\tilde{\bf \mu}_{k^{\prime}})}{T_{N}^{2}(T_{N}^{2} + \tilde{\bf f}^{2})} \right]^{2}
\end{equation}
where $w=\lambda^{2} B^{2}/(A_{e} D^{2} \Delta D)$, $D$ is the comoving distance to the center of the survey at redshift $z$, $\Delta D$ is the comoving depth of the survey, and $T_{\rm N}=\Delta V_{\rm N}({\bf U},\nu)$ as given by equation (\ref{eqn:noise}). This expression assumes that foregrounds are cleaned well below the signal, and that detector noise dominates over the signal, which is true for the first generation of instruments. The variance in a power spectrum estimate for a single $k$ mode, with line of sight component $k_{z} = \mu k$, is then: 
\begin{equation} 
\sigma_{\rm P}^{2}(k,\mu) = \left( {\bf F}^{{\bf k_{\perp}}^{-1}} \right)_{ii}. 
\end{equation}
Note that this effectively marginalizes over modes removed by foreground cleaning, since for such modes $\tilde{\bf \mu}_{k_{z}}^{\dagger} \tilde{\bf \mu}_{k_{z}^{\prime}} \rightarrow 0$, and hence $\sigma_{\rm P}^{2}(k,\mu) \rightarrow \infty$. The variance on the angle-averaged power spectrum over a spherical shell of logarithmic width $\epsilon = d \, {\rm ln} \, {\rm k}$ is then given by adding the error for individual $k$-modes in inverse quadrature:  
\begin{equation}
\frac{1}{\sigma_{\rm P}^{2}(k)}= \sum_{\mu} \frac{\epsilon k^{3} V_{\rm sur}}{4 \pi^{2}} \frac{\Delta \mu}{\sigma_{P}^{2}
(k,\mu)}
\end{equation}
where $V_{\rm sur}= D^{2} \Delta D (\lambda^{2}/A_{e})$ is the effective survey volume, and the sum over $\mu$ is over all modes in the upper half plane permitted by survey volume. The lower bound is set by the survey depth, while the upper bound is set by the telescope resolution. In practice, we do not sum over all Fourier cells in an annulus of constant $(k,\mu)$, but instead discretize k and physically count all modes within a given bin $k - \Delta k < |{\bf k}| < k + \Delta k$. This gives more accurate results in mode-counting, particularly at low wavenumber when the number of available modes is small. Note that the effective number of sampled modes is reduced at low $k$ after foreground cleaning (since modes with $k \mu \lsim k_{\rm z, crit}$ have $\sigma_{P}^{2}(k,\mu) \rightarrow \infty$), and so $\sigma^{2}_{\rm P}$ increases at low $k$, as was shown in Fig. \ref{fig:increased_variance}. 

We compare these Fisher matrix estimates against Monte-Carlo simulations. We first generate a Gaussian realization of MWA 21cm signal boxes, as described above. We also add telescope noise, as described in \S\ref{section:noise}, and foregrounds, as described in \S\ref{section:foregrounds}. We foreground clean with an $n=3$ polynomial, and then perform a maximum likelihood estimate of the power spectrum $P_{i}$ in 5 logarithmically spaced bins from $k=0.1$ to $k=0.7$, the approximate range of wavenumbers accessible by MWA. The Gaussian likelihood is:
\begin{equation}
{\cal L}(P_{i}|{\bf d}) = \frac{{\rm exp}(\frac{1}{2}{\bf d}^{T} {\bf C}^{-1} {\bf d})}{(2 \pi)^{N/2}({\rm det} \, {\bf C})^{1/2}}, 
\end{equation}
where ${\bf d}=\delta({\bf k})$ is the data in Fourier space, and the covariance matrix ${\bf C}={\bf S} + {\bf N}$ is assumed to be diagonal, where ${\bf S}({\bf k_{i},k_{j}}) = P_{\rm 21cm}(k_{i}) \delta_{ij}$, and the noise covariance matrix is: 
\begin{equation}
{\bf N} ({\bf k_{i},k_{j}}) = \left( \frac{\lambda^{2}}{A_{e}} \right)^{2} \frac{T_{\rm sys}^{2}}{B t_{o}} \frac{D^{2} \Delta D}{n (k_{\perp})} \delta_{ij}.
\end{equation}
In order to marginalize over modes which have been foreground cleaned, we set ${\bf N}_{ii}$ to a very large number, for modes where $\mu k_{i} < k_{\rm z, min}$. Since they are assigned a large variance, these modes are then given virtually no weight in the likelihood minimization. We set $k_{\rm z, min}$ to the first bin of $k_{z}$, as in the top panel of Fig.\ref{fig:k_z_cutoff}.\footnote{Note that using the first bin leaves us still with a very slight systematic bias, as is evident in the top panel of Fig.\ref{fig:k_z_cutoff}.
%which we removed by running many realization of Fig. \ref{fig:k_z_cutoff} and calculating a small correction factor $P_{\rm marg}/P_{\rm true}$. 
This is of course unnecessary and could have been avoided by a more judicious choice of $k_{\rm z,min}$.}. We perform the multi-dimensional minimization of the negative log likelihood, $f\equiv -2 {\rm ln} {\cal L}$, via a Neder-Simplex algorithm. We perform 200 Monte-Carlo simulations, and compute the mean and variance of our derived maximum-likelihood estimates, comparing them to the true input power spectrum and the Fisher matrix estimates for the variance respectively. 

\begin{figure}
%%%\epsscale{1.2}
\includegraphics[width=0.5\textwidth]{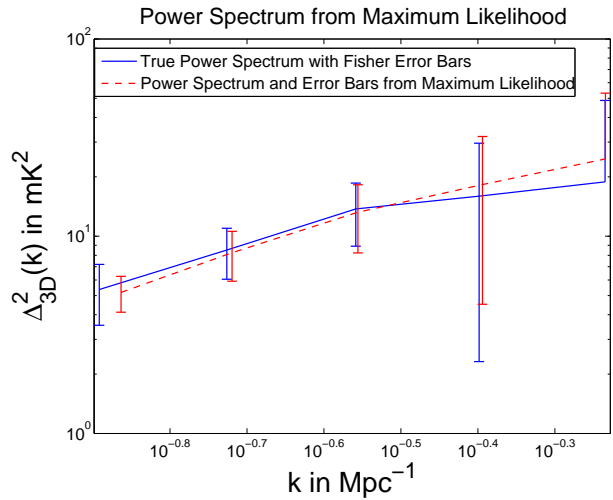}
\caption{The input power spectrum with error bars computed from the Fisher matrix, compared against the derived mean and standard deviation of the maximum likelihood solution of 200 Monte Carlo simulations of an MWA field. The two show excellent agreement. In particular, the fact that the mean derived power spectrum equals the true input power spectrum implies that our marginalization procedure enables unbiased estimates of the power spectrum, despite foreground cleaning, whilst the close correspondence with Fisher matrix error bars implies that we have a minimum variance estimator.}
\label{fig:max_likelihood}
\end{figure}

The results are shown in Fig. \ref{fig:max_likelihood}. The mean of the maximum likelihood estimates for the power spectrum shows excellent agreement with the input power spectrum. This implies that our marginalization procedure enables unbiased estimates of the power spectrum, despite the fact that unfettered foreground cleaning suppresses power across a broad range of scales. The variance of the power spectrum estimates also shows good agreement with Fisher matrix estimates, implying that we have a minimal variance estimator of the power spectrum which respects the Cramer-Rao bound. Note that we plot an asymmetric error bar on the final data point to avoid an error bar depicting negative power. Overall, mode marginalization is able to robustly handle the broad suppression of power due to foreground cleaning, which should therefore not be a problem. 

\section{Effects of Foreground Removal on Tomography}
\label{section:tomography}

\begin{figure}
%%%\epsscale{1.2}
\includegraphics[width=0.5\textwidth]{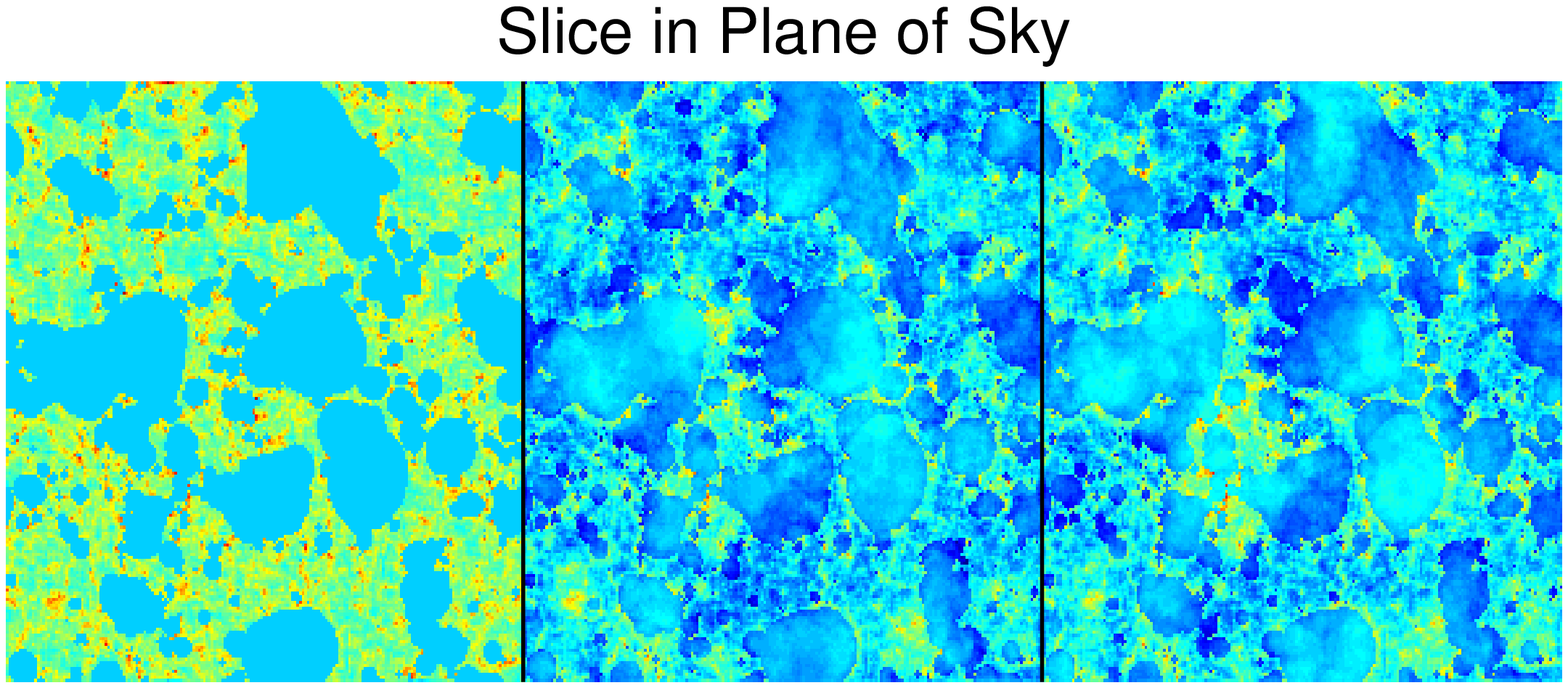} \\
\includegraphics[width=0.5\textwidth]{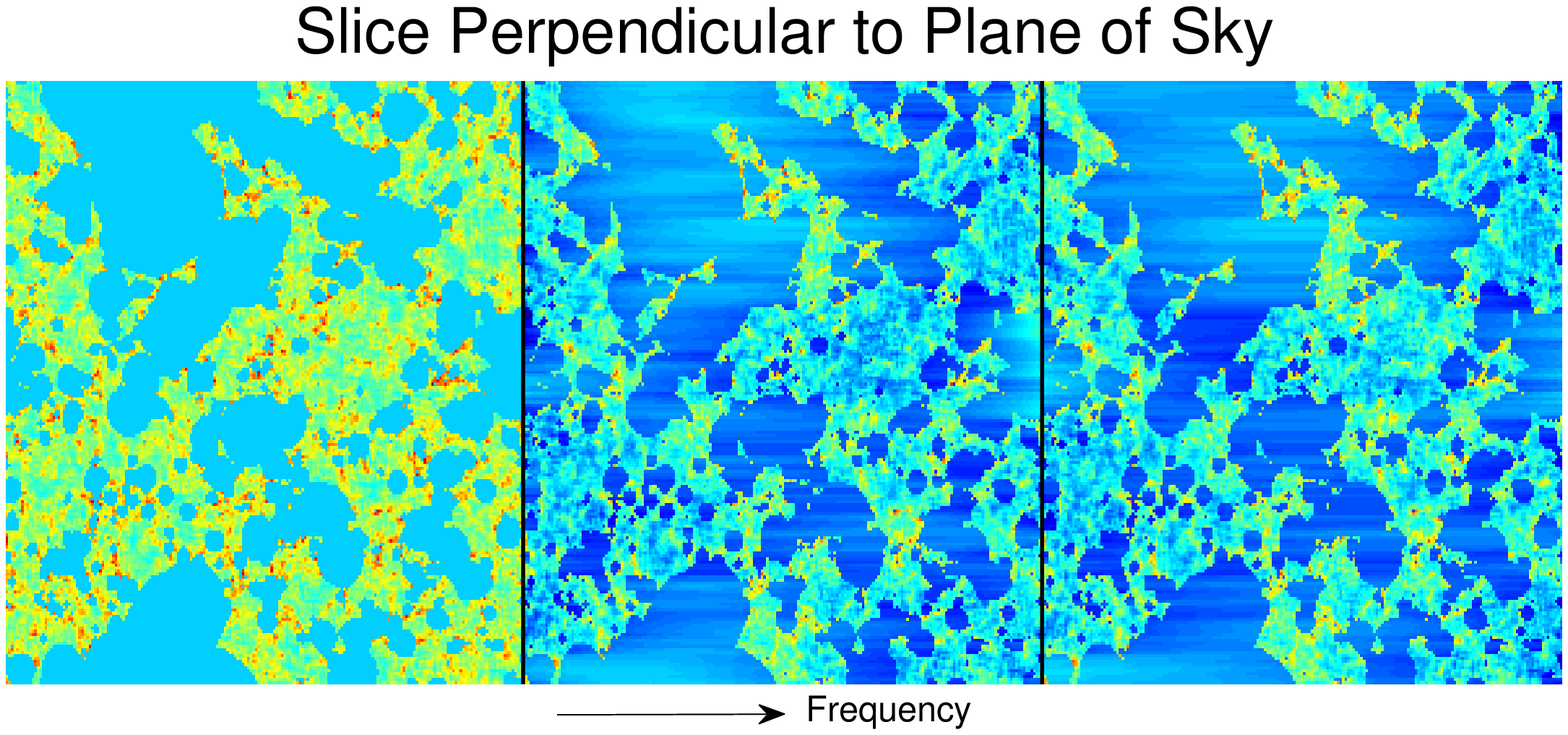}
\caption{Top panel shows the effect of foreground cleaning on a slice in the plane of the sky, while bottom panel shows the effect of foreground cleaning on a slice perpendicular to the plane of the sky. For both panels, left is true signal, center is $n=3$ polynomial fit, right is $n=2$ log fit. For a slice in the plane of the sky, the contrast between the bubbles and the neutral IGM appears to be reduced, although the topology of ionized regions is preserved. For a slice perpendicular to the plane of the sky, the reduction of contrast is also present, but to a lesser degree. Striae in the frequency direction become apparent.}
\label{fig:slice}
\end{figure}

3D tomography is precluded for the noise levels and angular resolution of the first generation of instruments, except for perhaps the exceptionally large bubbles blown by quasars; \citep{geil08a}. Nonetheless, the origin of the systematic artifacts introduced by foreground subtraction can perhaps best be understood in the image plane. We once again use the fact that polynomial fitting and subtraction can be represented by a linear projection operator (\S\ref{section:fg_removal}), and we therefore consider the impact of foreground cleaning on boxes of cosmological signal and foreground, without noise. Even in the presence of noise, the systematic artifacts due to the suppression of power in the cosmological signal will still be there (it will just be more difficult to pick out by eye). 

In Fig. \ref{fig:slice}, we show the effect of foreground cleaning on a slice of the box both parallel and perpendicular to the plane of the sky, for both a third order polynomial in $\nu$ and a second order polynomial in ${\rm log}(\nu)$ (we perform the latter to show that its effects are very similar to those of a polynomial in $\nu$. Of course, its use is deprecated since it cannot be represented as a linear projection operator). For a slice in the plane of the sky, the contrast between the bubbles and the neutral IGM appears to be reduced, although the topology of ionized regions is preserved. This reduction of contrast due to foreground cleaning will bedevil attempts to measure the 'step' in brightness temperature across an ionization front and hence potentially measure the neutral fraction, a fact which we previously pointed out in \citet{geil08}. For a slice perpendicular to the plane of the sky, the reduction of contrast is also present, but to a somewhat lesser degree. Instead, striae in the frequency direction become apparent. These are obviously due to inaccuracies in the fit to the foreground in the frequency direction. 

\begin{figure}
%%%\epsscale{1.2}
\includegraphics[width=0.5\textwidth]{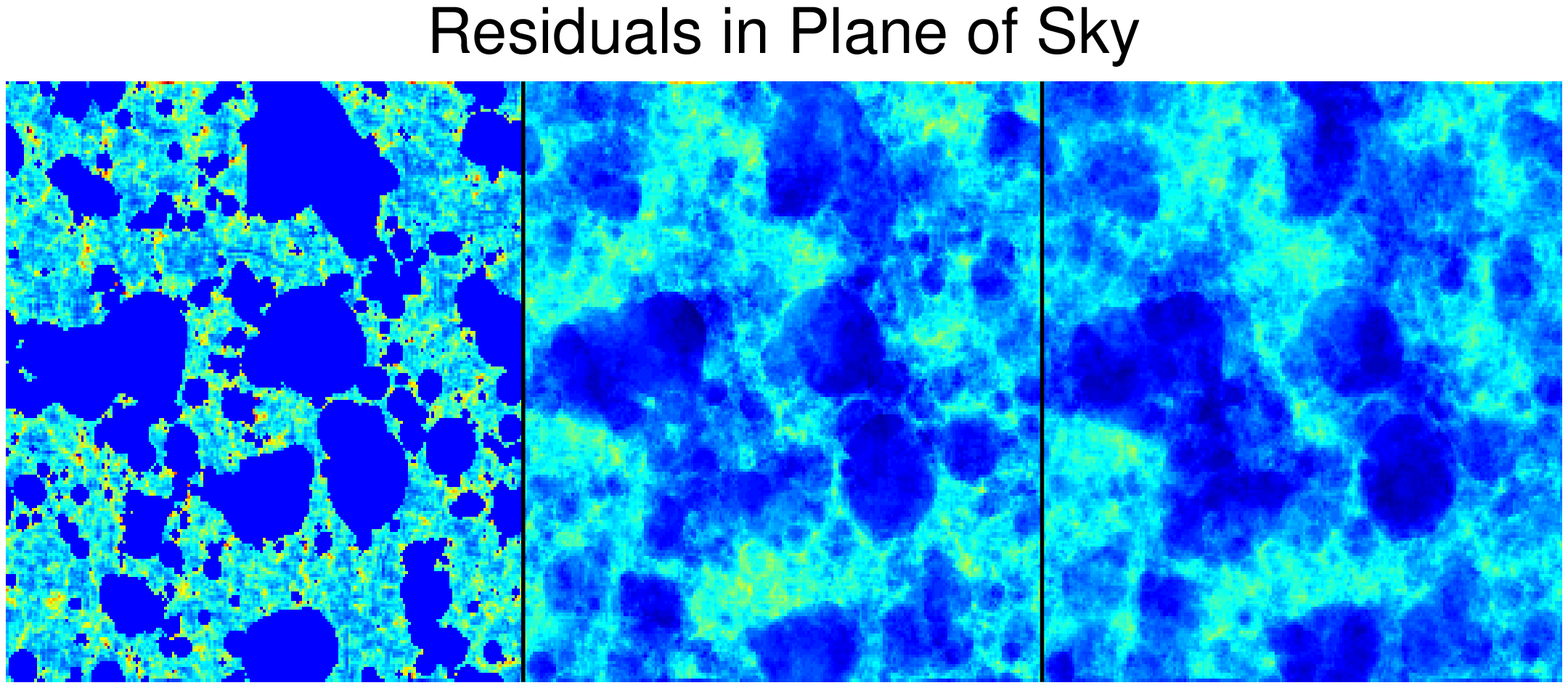} \\
\includegraphics[width=0.5\textwidth]{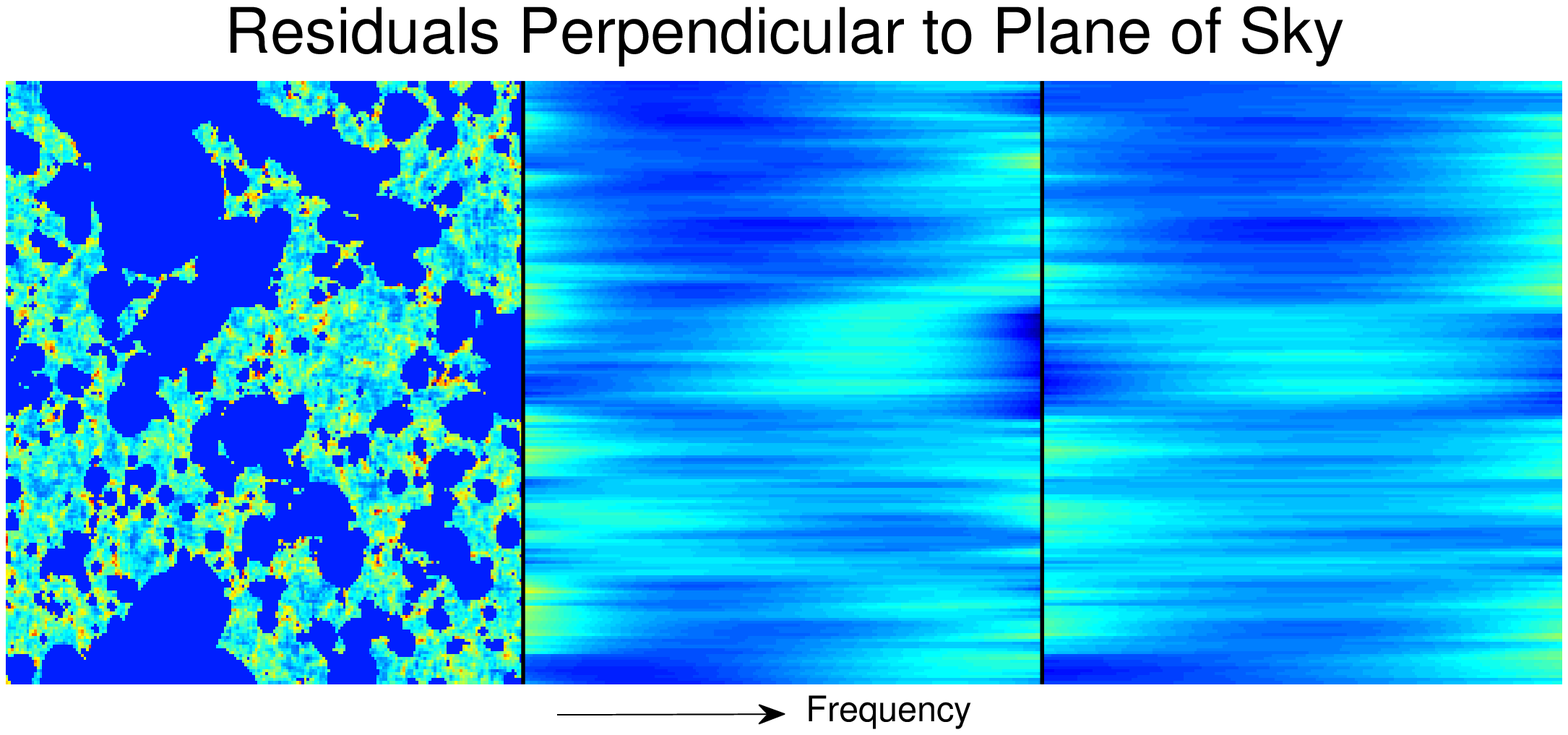}
\caption{Error in foreground subtraction ($\delta {\rm fg} = {\rm fg_{true}-fg_{fit}}$ in the plane of the sky (top panel) and perpendicular to the panel of the sky (bottom panel). As in Fig. \ref{fig:slice}, left is true signal, center is $n=3$ polynomial fit, right is $n=2$ log fit. In the plane of the sky, the errors in foreground subtraction are highly correlated with HII regions. Perpendicular to the plane of the sky, long wavelength errors in foreground subtraction create striation in the frequency direction, which is also correlated with large HII regions.}
\label{fig:residuals}
\end{figure}

\begin{figure}
\includegraphics[width=0.5\textwidth]{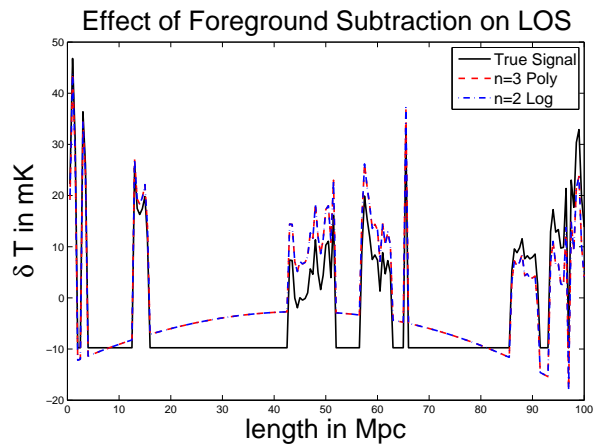}
\caption{The effect of foreground subtraction on a single line of sight. Errors in foreground subtraction are highly correlated with HII regions, which create downward curvature in the subtracted signal. }
\label{fig:los}
\end{figure}

These systematic artifacts are more apparent if we consider the residual errors in the foreground fit, or $\delta{\rm fg}={\rm fg_{true}}-{\rm fg_{fit}}$, which we show in Fig. \ref{fig:residuals}. From the top panel, which shows residuals in the plane of the sky, we see a striking result: {\it foreground fitting errors are not random, but instead highly correlated with the topology of ionized regions}. In fact, errors in foreground fitting trace out the distinctive structure of HII regions remarkably well. Note that $\delta{\rm fg}$ represents the difference between a (close to) power-law and a low order polynomial, so there is very little small scale power in the frequency direction, as can clearly be seen in the lower panel, which shows residuals perpendicular to the plane of the sky. However, closer examination of the bottom panel reveals that errors in foreground subtraction along the line of sight in fact also correlate with large scale HII regions. The foreground fit is 'pulled down' or 'pushed up' by large scale HII regions: ionized pixels are a source of highly correlated noise, since they tend to cluster together, and the polynomial fit to the foreground responds by curving incorrectly. We show an example of this along one line of sight in Fig. \ref{fig:los}. This small error in the foreground fit in the frequency direction occurs with sufficient consistency to produce the high correlated residuals we see in the plane of the sky in the top panel. Indeed, we see that that a substantial part of the signal has been subtracted off: the residuals show greater fidelity to the true signal (and the ionized regions have more correct contrast with the neutral regions) than the foreground subtracted box.

There are two lessons for us from this. Firstly, it is the presence of large HII regions which cause curvature in the foreground fit. In a sense, this is unsurprising, since it is the large HII regions which induce large scale power. If HII regions were small compared to the bandwidth we fit over, then there will be little distortion due to foreground fitting. Thus, as reionization progresses, and the typical bubble size increases, we need to fit over progressively larger and larger  bandwidths. We discuss this further in \S\ref{section:bandwidth}. Secondly, if we could identify the largest HII regions and use them as calibration points (we know that there is no 21cm signal there, so after foreground subtraction $\delta T_{b}=0$; in particular, all HII regions should be renormalized to the same level after foreground subtraction), this could greatly attenuate the distortions introduced by foreground cleaning. We discuss this possibility further in in \S\ref{section:bubbles}. 

\begin{figure}
\includegraphics[width=0.6\textwidth]{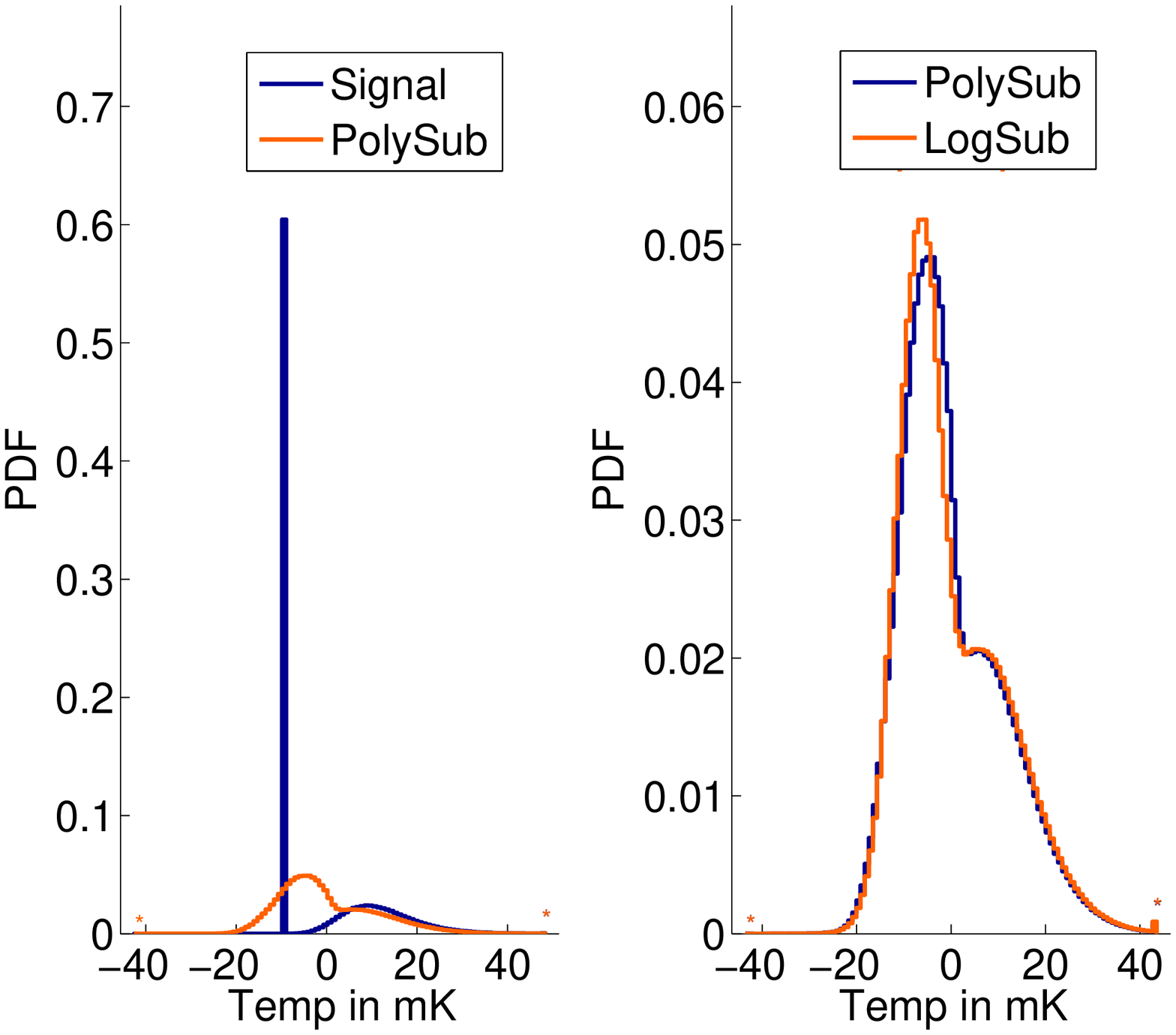}
\caption{Left Panel: The original signal PDF, and the PDF after foreground subtraction with an n=3 polynomial. Note that the mean temperature (which is not observable by an interferometer) has been subtracted from the signal PDF. Foreground cleaning considerable distortion of the PDF. Right panel: comparison of the foreground-cleaned PDFs for two different sets of basis functions. Both introduce comparable amounts of distortion.}
\label{fig:distorted_PDF}
\end{figure}

\section{Effect of Foreground Removal on PDF}
\label{section:PDF}

\begin{figure}
\includegraphics[width=0.5\textwidth]{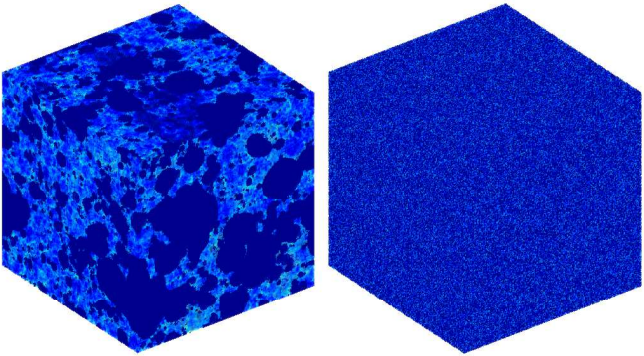}
\caption{Left: the original box of 21cm signal. Right: the scrambled box obtained by bootstrap resampling from the PDF of the original box. Such a box has an identical PDF, but a very different power spectrum.}
\label{fig:scrambled_box}
\end{figure}

\begin{figure}
\includegraphics[width=0.5\textwidth]{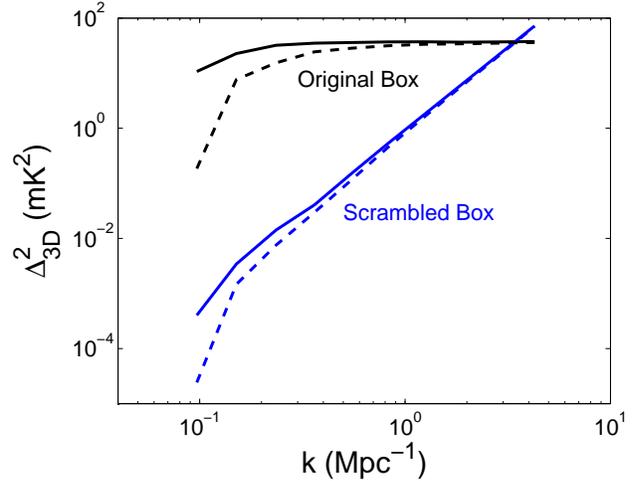}
\caption{Power spectrum of the original and scrambled box, both before (solid) and after (dashed) foreground subtraction. The scrambled box has a 'white noise' power spectrum $P(k)\approx$const$\Rightarrow \Delta^{2} \propto k^{3}$, which implies that it has much less large scale power, and hence is much less affected by foreground removal.}
\label{fig:ps_scrambled}
\end{figure}

\begin{figure}
\includegraphics[width=0.5\textwidth]{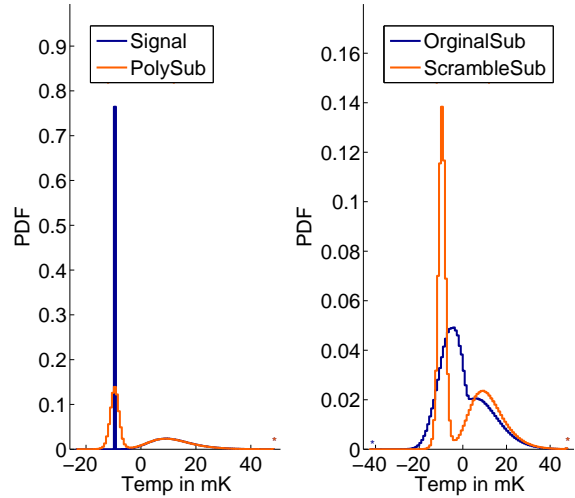}
\caption{Left panel: The original PDF of the scrambled box, and the PDF after foreground subtraction with an $n=3$ polynomial. Note that there is considerable less distortion of the PDF, compared to Fig \ref{fig:distorted_PDF}; in particular, the distribution of neutral pixels is almost perfectly preserved, whilst the distributed of ionized pixels is slightly broadened. Right panel: comparison of the foreground subtracted PDFs of the original (from Fig \ref{fig:distorted_PDF}) and scrambled boxes. The PDF of the scrambled box is significantly less distorted.}
\label{fig:pdf_scrambled}
\end{figure}

In principle, the 21cm PDF --- and in particular its redshift evolution --- contains a wealth of information about reionization \citep{furl04-21cmtop,wyithe07,harker09,ichikawa10}. In the left panel of Fig. \ref{fig:distorted_PDF}, we show our original signal PDF in blue, before foreground subtraction. A striking feature of the 21cm PDF is its bimodality: there is a delta function spike of ionized voxels, and a broad clump of neutral voxels, which is primarily due to density fluctuations. If one can measure the relative number of voxels in each distribution, it might be possible to measure the ionized fraction of the IGM. By doing this as a function of redshift, it would then be possible to chart the progress of reionization. In practice, this signal PDF will be convolved with telescope noise as well as the effects of finite telescope resolution (which then creates a long tail of partially ionized voxels). Since both of these effects are well understood, it is still possible to use maximum-likelihood mixture modeling to extract the cosmological signal (\citet{ichikawa10}; Oh et al, in preparation). This is despite substantial telescope noise; by eye, after the PDF appears perfectly Gaussian after it has been convolved with telescope noise (which is why we plot the noiseless distribution here). But the huge number of voxels means that small deviations from Gaussianity are detectable and have strong statistical significance. 

However, all this assumes that there are no systematic (rather than statistical) sources of noise. No work to date has consider the effects of foreground subtraction on the 21cm signal PDF; all have implicitly assumed perfect foreground subtraction. In fact, the distortions introduced by foreground cleaning are severe, and may preclude detailing modeling and statistical inferences from the measured PDF. The left panel of Fig. \ref{fig:distorted_PDF} shows the foreground subtracted signal PDF in green. The bimodality of the signal is clearly reduced, and the PDF is strongly distorted. In the right panel, we should that the distorted PDFs which emerge from cleaning with a third order polynomial in $\nu$ or a second-order polynomial in ${\rm log} \, \nu$ are comparable. In retrospect, these strong distortions could have been anticipated by the results of \S\ref{section:tomography}, where we saw the sharp reduction in contrast between neutral and ionized regions.

One could imagine at least two possible sources of this distortion. One is that least-squares regression is an optimal foreground estimator when noise is Gaussian. In this case, the 'noise', or the 21cm signal we are trying to recover, is highly non-Gaussian, and perhaps least-squares regression breaks down (as it does when there are significant outliers in the data). Least-squares regression might be trying to find the solution which has maximally Gaussian residuals, and thus destroys the very non-Gaussian signature we seek. In this case, we would want to use robust regression, minimizing some functional other than the sum of squares. Alternatively, the distortion could be due to the removal of large scale power, as we have seen previously. This would require a different damage control strategy. 

To distinguish between these possibilities, we create a new signal box. We bootstrap resample (with replacement) from the original signal box, but impose no spatial correlations between voxels. The result is the box on the right panel of Fig. \ref{fig:scrambled_box}, which clearly has no large scale features, but by construction the same PDF as the original signal box. As can be seen in Fig. \ref{fig:ps_scrambled}, the scrambled box has a white noise power spectrum, $P(k)\approx$const$\Rightarrow \Delta^{2} \propto k^{3}$, which implies that it has much less large scale power, and hence the fluctuation spectrum is much less affected by foreground removal. Indeed, the rms temperature fluctuations (which are dominated by small scales) are scarcely affected by foreground removal, unlike the original signal box. The left panel of Fig. \ref{fig:pdf_scrambled} compares the PDF of the original and foreground subtracted boxes; notice that the PDF is only slightly distorted. In particular, the distribution of neutral pixels is almost perfectly preserved, whilst the distributed of ionized pixels is only slightly broadened; the bimodality of the pixel distribution is still strongly apparent. In the right panel, we overlay the foreground subtracted PDFs of the original (from Fig \ref{fig:distorted_PDF}) and scrambled boxes. The PDF of the scrambled box is significantly less distorted. This shows that the distortion of the PDF during foreground removal must be due to the suppression of large power, since only the original box has significant large scale power. Non-Gaussianity must play only a small role, since both boxes have identical non-Gaussian PDFs. 

In summary, foreground removal causes significant distortion of the PDF, and it is directly linked to the removal of large scale power from the signal. Any detailed modeling of the 21cm PDF must find a way to overcome this; otherwise, inferences about reionization will be systematically biased. Furthermore, the degree of distortion will be redshift dependent, increasing as reionization progresses, and the characteristic bubbles size (and hence amount of large scale power) increases. 

\section{Possible Improvements to Foreground Removal Methods}
\label{section:improvements}

\begin{figure}
\includegraphics[width=0.5\textwidth]{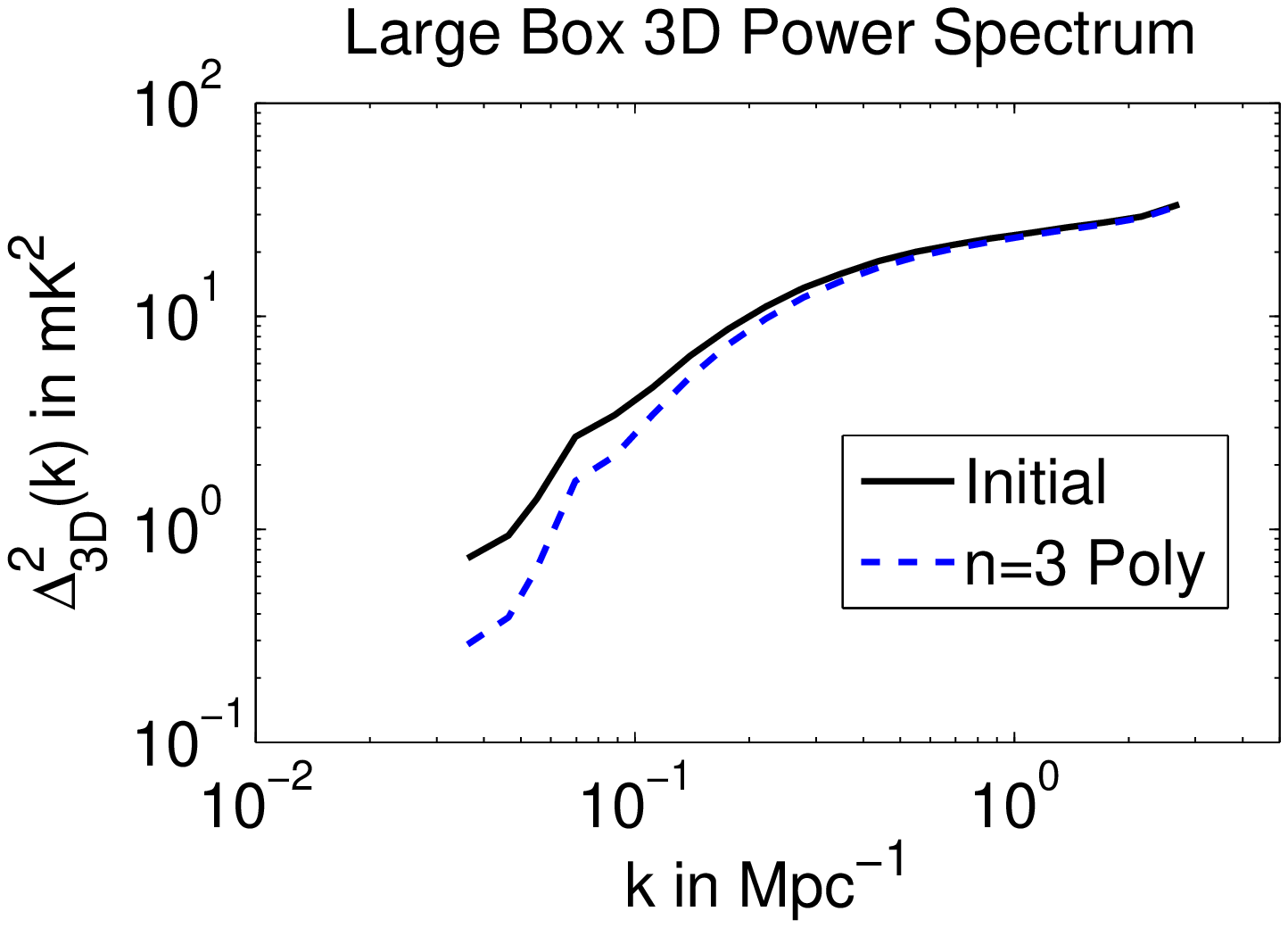}
\includegraphics[width=0.5\textwidth]{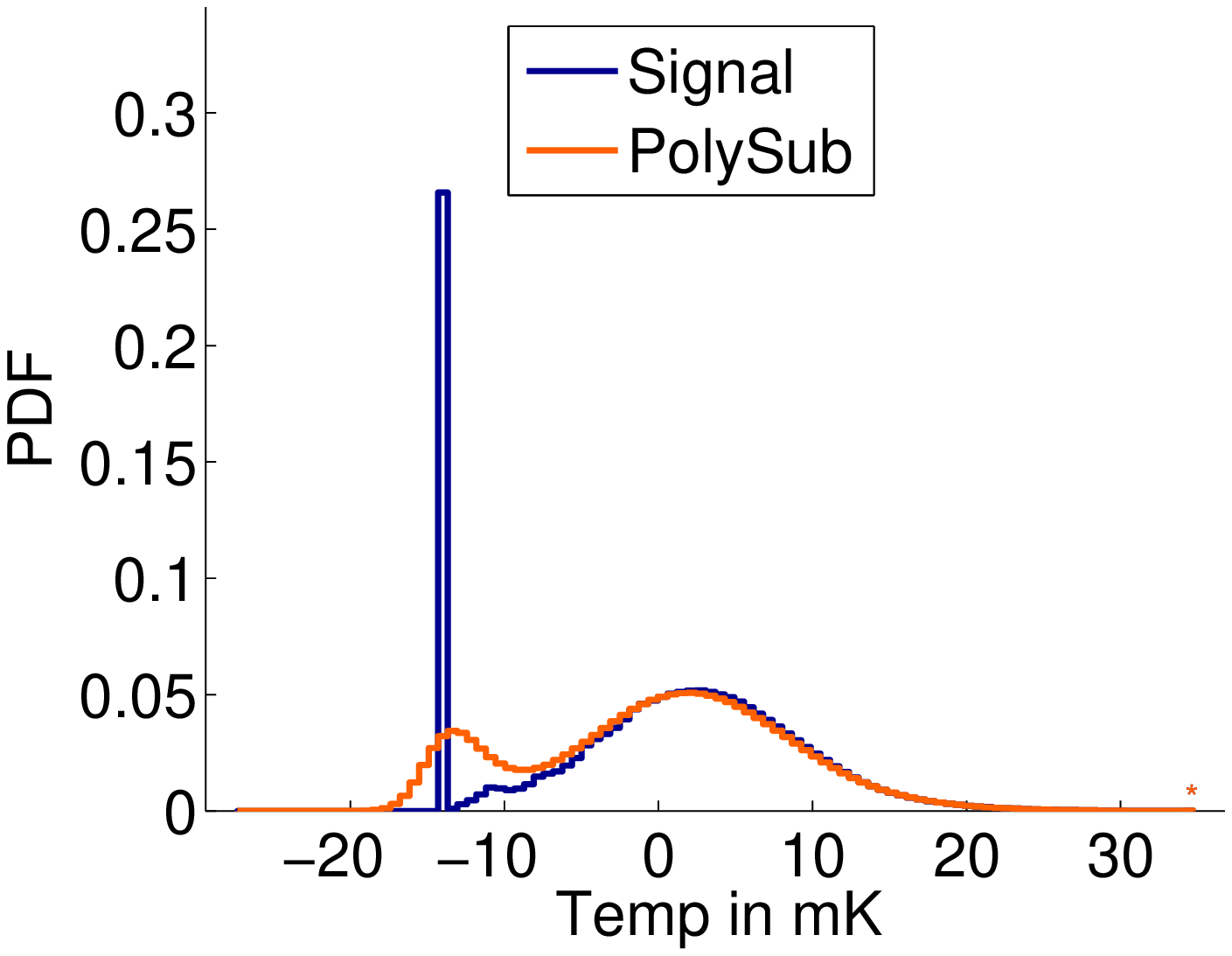}
\caption{Foreground subtraction in a $400^{3} \, ({\rm cMpc})^{3}$ box where $x_{\rm HI}=0.3$. Here, the characteristic bubble size is much smaller than the comoving width of the slice, and hence there is little fluctuation power on the large scales over which foreground subtraction suppresses power (top panel).  Hence, as in the 'scrambled box' of \S\ref{section:PDF}, there comparatively little distortion of the 21cm PDF (bottom panel); in particular, the distribution of neutral pixels is almost perfectly preserved.} 
\label{fig:big_box}
\end{figure}

We have seen that distortion of the measured power spectrum can by treated by marginalizing over the appropriate modes with $k_{z} \lsim k_{\rm z,min}$; we can then recover an unbiased estimator of the power spectrum, albeit with increasing variance at small wavenumber, as shown in Fig. \ref{fig:increased_variance}. On the other hand, there is no analogous marginalization procedure which can recover the undistorted PDF and (in the future) 21cm tomographic images. Here, we suggest two ways in which these distortions can potentially be overcome, which can be applied at the early and late stages of reionization respectively. This is fertile ground for future work. 

\subsection{Maximizing the ratio of Bandwidth to Bubble Size} 
\label{section:bandwidth}

Figures \ref{fig:ps_scrambled} and \ref{fig:pdf_scrambled} show that if the signal has relatively little large scale power to begin with, then foreground subtraction will produce little distortion of the PDF. Large scale power is generally a function of HII bubble size; the larger the bubbles are, the more large scale power there will be. The characteristic bubble size increases continually throughout reionization; there is less large-scale power during the early stages of reionization. Thus, PDF distortion should be less severe earlier on. On the other hand, Fig. \ref{fig:window} shows that foreground cleaning suppresses power on the characteristic lengthscale of the frequency slice over which the cleaning is done; if we can clean over larger lengthscales, then the modes we are interested in may be preserved (\citet{mcquinn06} have also made the latter point). 

These considerations suggest that PDF distortion will be smaller if we maximize the ratio of bandwidth to bubble size. Thus, the earlier stages of reionization will be less susceptible to PDF distortion, and we should use as large a bandwidth as possible to do the cleaning\footnote{Note that it is not optimal to analyze the PDF over large bandwidths when cosmic evolution can be important. Rather, one should clean the data over larger frequency slices, and then analyze it over narrower slices, within which the neutral fraction is relatively constant. Over a broader bandwidth, the foreground shows greater curvature, and eventually one cannot clean the foreground well below the signal with a given polynomial fit.}. Fig. \ref{fig:big_box} shows the power spectrum and the PDF of a $400^{3} \, ({\rm cMpc})^{3}$ box at $z=8$ with $x_{\rm HI}=0.3$ (note that out fiducial $100^{3} \, ({\rm cMpc})^{3}$ box at $x=8$ has $x_{\rm HI}=0.56$). As expected, the power spectrum has little power on large scales; thus, there is indeed little distortion of the PDF due to foreground cleaning. We can likely measure the PDF fairly accurately during the earlier stages of reionization; however, during the later stages, we must use a different strategy.

\subsection{Large HII regions as Foreground Calibrators}
\label{section:bubbles}

\begin{figure}
\includegraphics[width=0.49\textwidth]{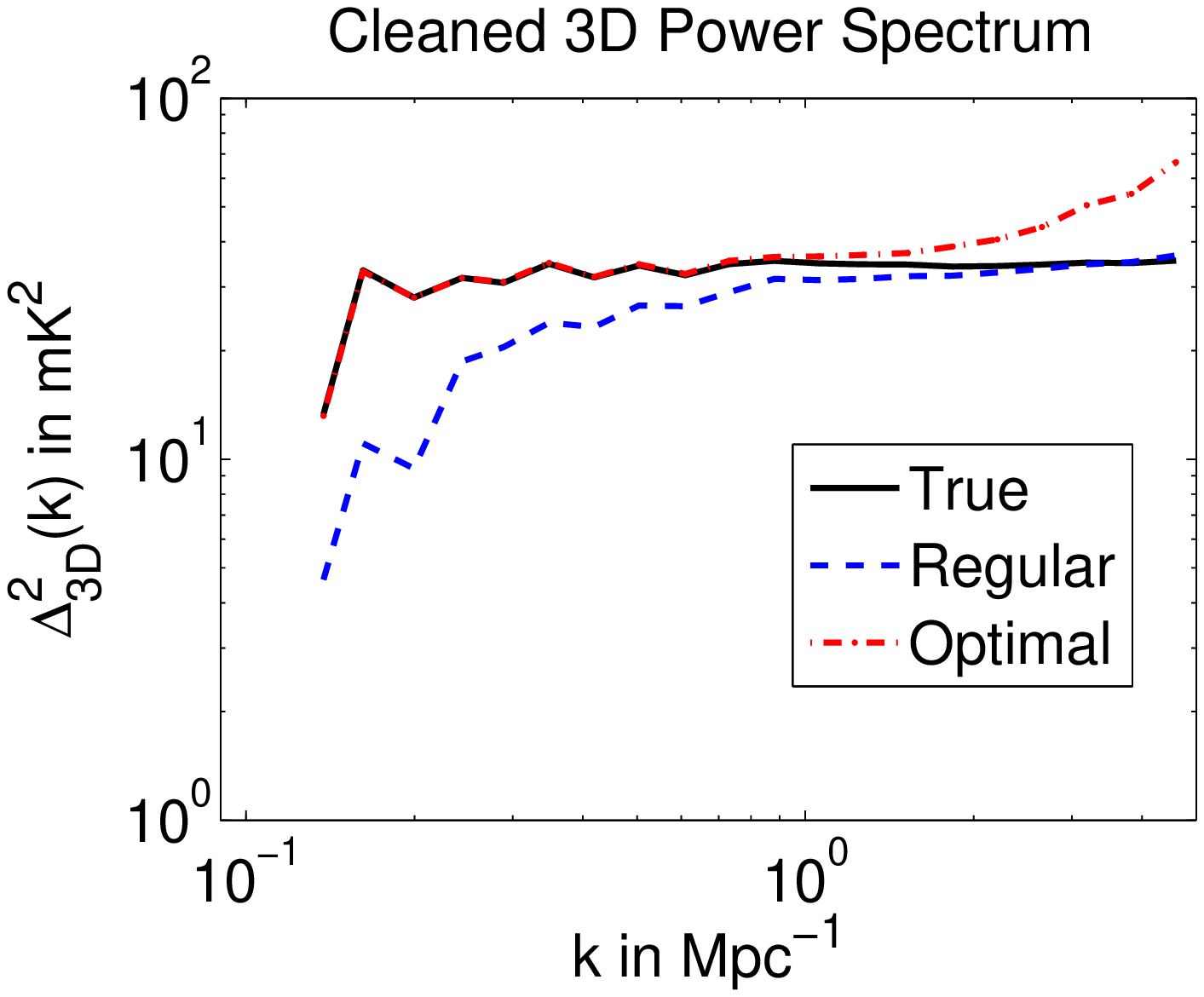}
\includegraphics[width=0.49\textwidth]{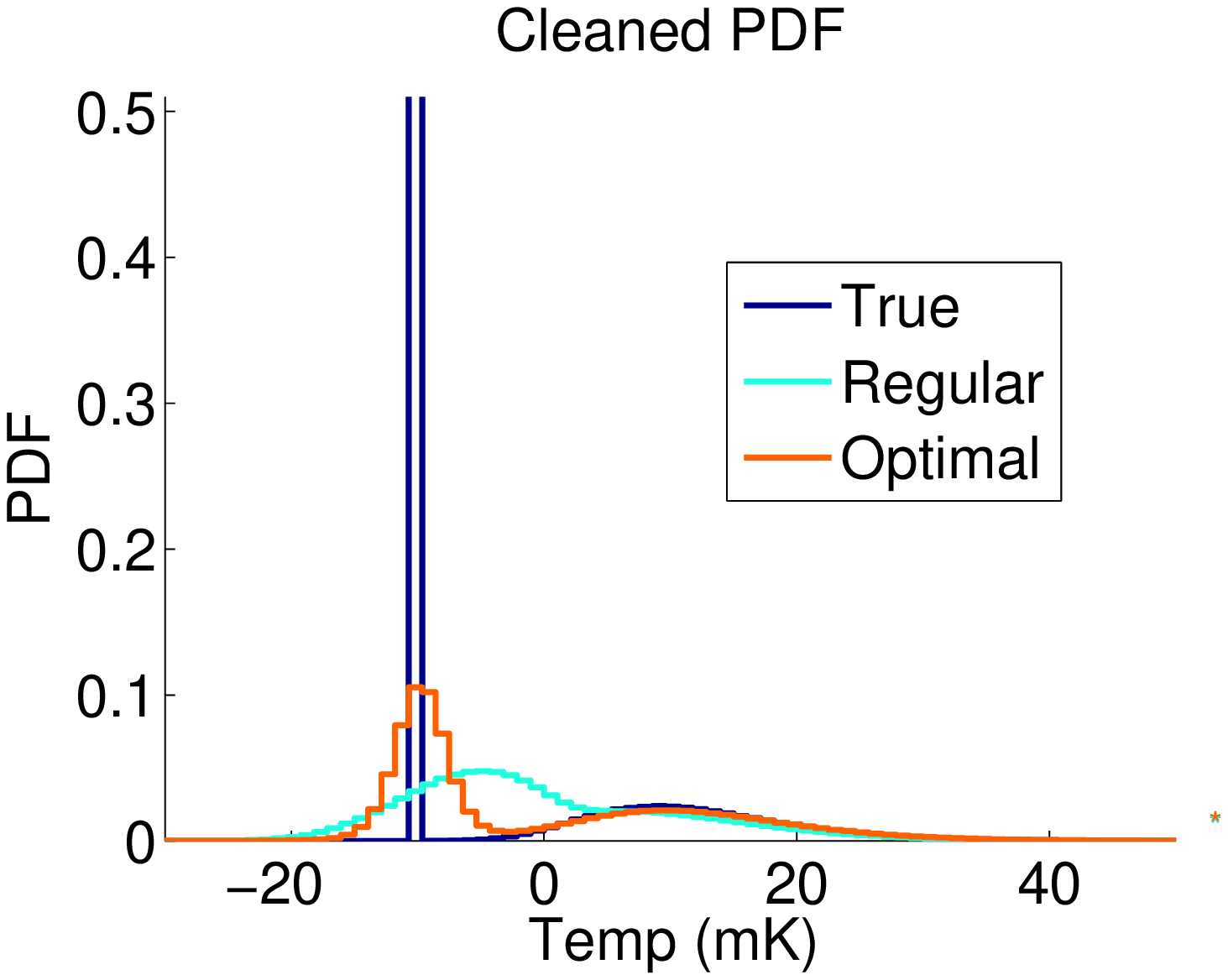}
\includegraphics[width=0.49\textwidth]{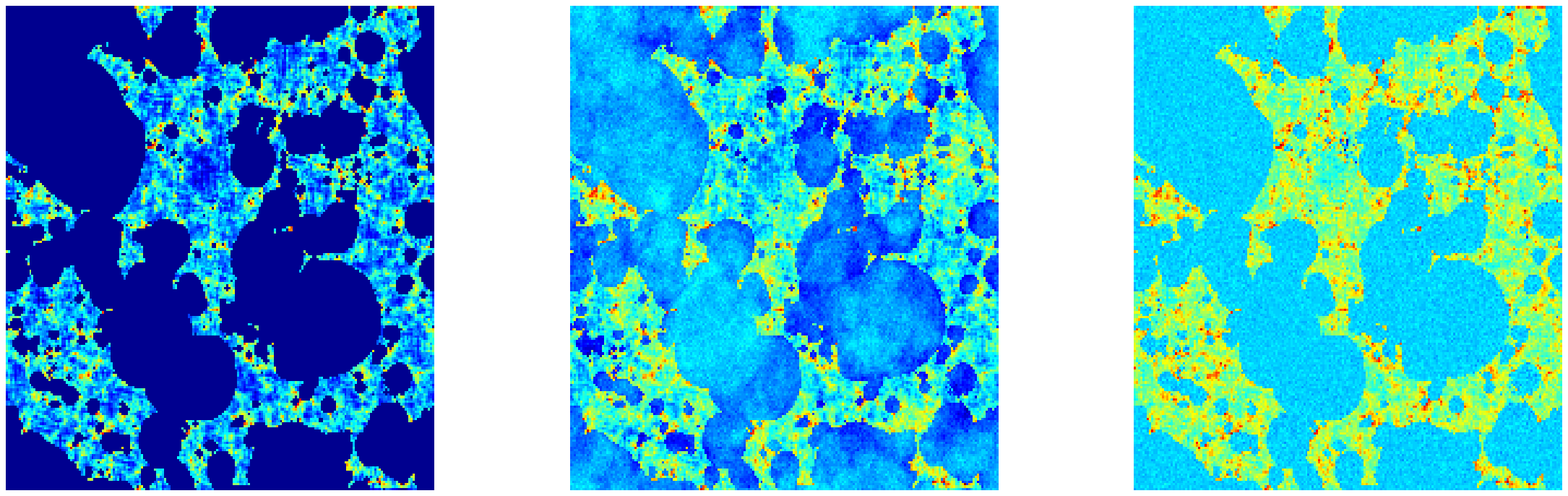}
\caption{Comparison of the customary foreground cleaning method ('Regular') and cleaning with an additional step when the positions of bubbles are known ('Optimal'), for our usual $100^{3} \, ({\rm cMpc})^{3}$ box.  Foregrounds are fit in the presence of noise, with S/N$\sim 1$. Top panel: large scale power is correctly recovered with the optimal algorithm. The bias at small scales is due to errors in the foreground fit due to noise. Middle panel: the distortion of the PDF is far less severe with the optimal algorithm, since large scale power is correctly preserved. Bottom panel: a slice in the plane of the sky of the original box (left), regularly cleaned (middle), and optimally cleaned box (right). Large bubbles have all been set to the same zero-point in the right panel, compared to the varied contrast in the middle panel.} 
\label{fig:optimal_estimator}
\end{figure}

In \S\ref{section:tomography}, we saw that large HII bubbles were the primary reason for errors in foreground fitting. The foreground fit is Õpulled downÕ or Õpushed upÕ by large scale HII regions: ionized pixels are a source of highly correlated noise, since they tend to cluster together, and the polynomial fit to the foreground responds by incorrect curvature. Examination of Fig. \ref{fig:los} suggests a solution: if we could 'renormalize' the ionized HII regions all to the same level (since there is no 21cm signal there), then the incorrect curvature would vanish, and the large scale power incorrectly subtracted from the box would be restored. In practice, this can be accomplished by a second stage after the initial round of foreground cleaning: fitting a polynomial {\it only to the voxels in ionized regions}. Subtracting off this polynomial ensures that all HII regions will be 'renormalized' to the same zero-point, as required.  

Applying this to a noiseless box with foregrounds results in {\it perfect} recovery of the original signal box! Of course, in a realistic scenario, there are two problems: (i) we will not know a priori the positions of fully ionized voxels; furthermore, finite telescope resolution will 'mix' neutral and ionized voxels, producing voxels with effective partial ionization, and reducing the number of fully ionized calibration points. (ii) We fit to foregrounds in the presence of noise, and so even if we knew where all the ionized voxels were, they scatter about the true zero-point. This scatter produces fitting errors in the second cleaning stage. Regarding (i), note that even if direct imaging of bubbles is difficult in the first generation of instruments, it is possible to classify voxels as belonging to ionized or neutral regions in a probabilistic fashion, via mixture model or tesselation algorithms. Image segmentation and/or edge detection in the presence of noise is a classic image processing problem, and there are many possible methods (e.g., Canny algorithm, Voronoi teselation, void-finding algorithms (e.g., see \citet{colberg08}), etc). Note that the impact of both (i) and (ii) are reduced when bubbles are large, in the later stages of reionization: the HII regions are easier to detect, and also the noise is averaged over more voxels. Fortunately, this is the regime in which this second cleaning step becomes necessary, since large bubbles are the ones to destabilize the foreground fit. 

We leave detailed quantitative investigation of such a cleaning algorithm in a realistic setting in conjunction with an image segmentation algorithm to future work. Here, we simply illustrate its potential. We consider an idealized case where we know the positions of all ionized voxels (in fact, only knowing the voxels in large HII regions gives comparable results), and add Gaussian noise such that $S/N \sim 1$ for each voxel; this is less noise than is realistic for a first-generation instrument. We then fit for foregrounds in the presence of noise, performing the second 'renormalization' step as described. We subtract the imperfectly fitted foreground from a box with signal and foreground only, so that any errors in foreground subtraction are evident. The results are in Fig. \ref{fig:optimal_estimator}. We see from the top panel that the large scale power spectrum is correctly recovered with the optimal algorithm. Thus, using this cleaning procedure, we may not have to marginalize over modes with low $k_{\rm z}$, and pay the price in increased sample variance at low $k$. The bias at high $k$ is due to errors in the foreground fit due to noise; in practice, these will be correctly handled when solving via a maximum likelihood scheme.  In the middle panel, we see that the distortion of the PDF is far less severe with the optimal algorithm, and the intrinsic bimodality more correctly preserved. In the bottom panel, we compare a slice in the plane of the sky of the original box (left), a box cleaned with the regular algorithm (middle), and a box cleaned with the optimal algorithm (right). Note that large bubbles have all been set to the same zero-point in the right panel, compared to the varied contrast in the middle panel. This directly results in the reduced distortion of the PDF. 

\section{Conclusions}
\label{section:conclusions}

In this paper, we focus on the removal of large scale power along the line of sight by foreground cleaning algorithms for upcoming 21cm surveys of the high-redshift universe. Whilst the subtraction of large scale power is well-known, the systematic biases it induces in the cleaned data, and how this propagates into various statistics, has not been well-understood. We emphasize that there are many aspects of the foreground removal problem that we do not address. For instance, we do not consider 'mode-mixing' and the consequences of bright source subtraction (\citetalias{liu09a,liu09b}; \citet{bowman09,datta10}); our foreground model is much less sophisticated than others (e.g., \citet{jelic08}); we largely ignore instrumental effects, such as leakage of polarized foregrounds to the total signal \citep{jelic10,geil10}; and we do not examine the subtleties of various foreground cleaning algorithms (e.g., \citep{harker09}). %For instance, we do not consider the frequency dependence of beam, which can introduce frequency ripples ('mode mixing'); calibration errors; residual contamination due to position errors in the sky model for bright sources; the relative efficacy of cleaning in real space vs. Fourier space; polarization of foregrounds; or non-parametric foreground subtraction schemes. 
Such issues have been dealt with by other authors. However, the removal of large scale power along the line of sight is to date common to all foreground subtraction schemes. We specialize to the case of linear subtraction schemes, and use the fact that the impact of foreground cleaning on the signal can therefore be considered in isolation, without considering its interaction with noise. 

Our principal conclusions are as follows: 
\begin{itemize}
\item{Removal of large scale power along the line of sight aliases into suppression of power across a broad range of scales in 3D, an effect which can be understood analytically. We compare analytic expectations to foreground cleaning simulations and find excellent agreement. This systematic underestimate of the true power spectrum can be correctly accounted for by marginalizing over all modes with $k_{z} \leq k_{\rm z, crit}$. However, the reduction in the number of measurable modes at a given wavenumber unavoidably increases sample variance.}
\item{We show that foreground cleaning has the same effect on Gaussian realizations of the power spectrum as it does on fully non-Gaussian boxes. This allows for rapid Monte-Carlo simulations. We perform maximum-likelihood Monte-Carlo simulations of power spectrum estimation with realistic noise, where we correctly marginalize over foreground cleaned modes, and show that we are able to recover unbiased power spectrum estimates, with the minimum variance given by Fisher matrix estimates.}
\item{To date there have been no studies of the impact of foreground cleaning on the PDF. We show that in fact significant distortion of the PDF occurs, which would compromise attempts to mine the 21cm PDF for astrophysical information. By comparing two boxes with the same non-Gaussian PDF but very different power spectra, we show that the distortion is due to the removal of large scale power, rather than errors in least-squares regression due to the non-Gaussian PDF.}
\item{Foreground cleaning will also distort tomographic images, by reducing the contrast between neutral and ionized regions. However, the topology of ionized regions is preserved. The reduction in contrast is because errors in foreground cleaning strongly correlate with the presence of ionized regions, which lead to artificial curvature in the foreground fit.}
\item{Whilst the effect of foreground cleaning on the power spectrum is easily dealt with, correcting distortions in the PDF and tomographic images is more difficult. We make two suggestions. The first is to foreground clean over the largest feasible bandwidth over which the effects of evolution can be neglected. Since foreground cleaning removes power on the scale of the cleaning bandwidth, if there is little power on these scales, there will be little distortion of the PDF or reduction of contrast between neutral and ionized regions. Essentially, the slice width must be significantly larger than the characteristic bubble size. Early in reionization, this constraint will be easy, but more difficult in the late stages, as bubbles grow. For these later stages, identification of the largest ionized regions (which consist of foreground emission only) provides calibration points which potentially allow recovery of large-scale modes. Detailed exploration of this is an obvious avenue for future work.}
\item{In the Appendices, we put to rest two concerns which have been raised from time to time, but never been calculated in detail. One is that because of the extremely bright nature of continuum foregrounds, even very slight deviations from spectral smoothness could introduce frequency structure which would swamp the 21cm signal. We show that synchrotron and free-free emission for even a single electron has such a broad frequency response that temperature fluctuations over the small measured frequency intervals are negligible; averaging over a huge number of electrons of course damps any fluctuations even further. Another concern is that the integrated extragalactic radio recombination line background might constitute a formidable foreground. We show that even with aggressive estimates for stimulated emission from star-forming galaxies---where RRL luminosity is $\sim 10\%$ of the radio continuum luminosity--the RRL background is unlikely to prove a problem. The reason is the small comoving volume probed: whereas all continuum sources along a line of sight contribute, for a specific recombination line, RRL emitters from only a tiny redshift slice contribute.} 
\end{itemize}

\section*{Acknowledgments}
We thank Steve Furlanetto, Paul Geil, Adam Lidz, Matt McQuinn for helpful conversations and email correspondence. We are also grateful to Tzu-Ching Chang for early discussions on foreground removal, and Andrei Mesinger for the semi-numeric 21cm box used here. We acknowledge NSF grant AST 0908480 for support. SPO thanks the Aspen Center for Physics and UCLA for hospitality during the completion of this paper. 

\bibliography{master_references} 

\begin{thebibliography}{}

\bibitem[\protect\citeauthoryear{{Anantharamaiah}, {Viallefond}, {Mohan},
  {Goss} \& {Zhao}}{{Anantharamaiah} et~al.}{2000}]{anantharamaiah00}
{Anantharamaiah} K.~R.,  {Viallefond} F.,  {Mohan} N.~R.,  {Goss} W.~M.,
  {Zhao} J.~H.,  2000, \apj, 537, 613

\bibitem[\protect\citeauthoryear{{Anantharamaiah}, {Zhao}, {Goss} \&
  {Viallefond}}{{Anantharamaiah} et~al.}{1993}]{anantharamaiah93}
{Anantharamaiah} K.~R.,  {Zhao} J.-H.,  {Goss} W.~M.,    {Viallefond} F.,
  1993, \apj, 419, 585

\bibitem[\protect\citeauthoryear{{Banday} \& {Wolfendale}}{{Banday} \&
  {Wolfendale}}{1991}]{banday91}
{Banday} A.~J.,  {Wolfendale} A.~W.,  1991, \mnras, 248, 705

\bibitem[\protect\citeauthoryear{{Boissier}}{{Boissier}}{2005}]{boissier05}
{Boissier} S. e.~a.,  2005, \apjl, 619, L83

\bibitem[\protect\citeauthoryear{{Bond}, {Cole}, {Efstathiou} \&
  {Kaiser}}{{Bond} et~al.}{1991}]{bond91}
{Bond} J.~R.,  {Cole} S.,  {Efstathiou} G.,    {Kaiser} N.,  1991, \apj, 379,
  440

\bibitem[\protect\citeauthoryear{{Bowman}, {Morales} \& {Hewitt}}{{Bowman}
  et~al.}{2008}]{bowman-foreground-08}
{Bowman} J.~D.,  {Morales} M.~F.,    {Hewitt} J.~N.,  2008, ArXiv e-prints

\bibitem[\protect\citeauthoryear{{Bowman}, {Morales} \& {Hewitt}}{{Bowman}
  et~al.}{2009}]{bowman09}
{Bowman} J.~D.,  {Morales} M.~F.,    {Hewitt} J.~N.,  2009, \apj, 695, 183

\bibitem[\protect\citeauthoryear{{Colberg} J.~M.}{{Colberg}}{2008}]{colberg08}
{Colberg} J.~M. e.,  2008, \mnras, 387, 933

\bibitem[\protect\citeauthoryear{{Datta}, {Bowman} \& {Carilli}}{{Datta}
  et~al.}{2010}]{datta10}
{Datta} A.,  {Bowman} J.~D.,    {Carilli} C.~L.,  2010, ArXiv e-prints

\bibitem[\protect\citeauthoryear{{Di Matteo}, {Ciardi} \& {Miniati}}{{Di
  Matteo} et~al.}{2004}]{dimatteo04}
{Di Matteo} T.,  {Ciardi} B.,    {Miniati} F.,  2004, \mnras, 355, 1053

\bibitem[\protect\citeauthoryear{{Di Matteo}, {Perna}, {Abel} \& {Rees}}{{Di
  Matteo} et~al.}{2002}]{dimatteo02}
{Di Matteo} T.,  {Perna} R.,  {Abel} T.,    {Rees} M.~J.,  2002, \apj, 564, 576

\bibitem[\protect\citeauthoryear{{Furlanetto}, {Oh} \& {Briggs}}{{Furlanetto}
  et~al.}{2006}]{furl06-review}
{Furlanetto} S.~R.,  {Oh} S.~P.,    {Briggs} F.~H.,  2006, \physrep, 433, 181

\bibitem[\protect\citeauthoryear{{Furlanetto}, {Zaldarriaga} \&
  {Hernquist}}{{Furlanetto} et~al.}{2004}]{furl04-21cmtop}
{Furlanetto} S.~R.,  {Zaldarriaga} M.,    {Hernquist} L.,  2004, \apj, 613, 16

\bibitem[\protect\citeauthoryear{{Geil}, {Gaensler} \& {Wyithe}}{{Geil}
  et~al.}{2010}]{geil10}
{Geil} P.~M.,  {Gaensler} B.~M.,    {Wyithe} J.~S.~B.,  2010, ArXiv e-prints

\bibitem[\protect\citeauthoryear{{Geil} \& {Wyithe}}{{Geil} \&
  {Wyithe}}{2008}]{geil08a}
{Geil} P.~M.,  {Wyithe} J.~S.~B.,  2008, \mnras, 386, 1683

\bibitem[\protect\citeauthoryear{{Geil}, {Wyithe}, {Petrovic} \& {Oh}}{{Geil}
  et~al.}{2008}]{geil08}
{Geil} P.~M.,  {Wyithe} J.~S.~B.,  {Petrovic} N.,    {Oh} S.~P.,  2008, \mnras,
  390, 1496

\bibitem[\protect\citeauthoryear{{Gleser}, {Nusser} \& {Benson}}{{Gleser}
  et~al.}{2008}]{gleser08}
{Gleser} L.,  {Nusser} A.,    {Benson} A.~J.,  2008, \mnras, 391, 383

\bibitem[\protect\citeauthoryear{{Gnedin} \& {Shaver}}{{Gnedin} \&
  {Shaver}}{2004}]{gnedin04}
{Gnedin} N.~Y.,  {Shaver} P.~A.,  2004, \apj, 608, 611

\bibitem[\protect\citeauthoryear{{Gordon} \& {Sorochenko}}{{Gordon} \&
  {Sorochenko}}{2002}]{gordon_rrl_book}
{Gordon} M.~A.,  {Sorochenko} R.~L.,  2002, {Radio Recombination Lines. Their
  Physics and Astronomical Applications}.
Astrophysics and Space Science Library, Vol.~282.~Kluwer Academic Publishers,
  Dordrecht

\bibitem[\protect\citeauthoryear{{Harker}, {Zaroubi}, {Bernardi}, {Brentjens},
  {de Bruyn}, {Ciardi}, {Jeli{\'c}}, {Koopmans}, {Labropoulos}, {Mellema},
  {Offringa}, {Pandey}, {Pawlik}, {Schaye}, {Thomas} \& {Yatawatta}}{{Harker}
  et~al.}{2010}]{harker10}
{Harker} G.,  {Zaroubi} S.,  {Bernardi} G.,  {Brentjens} M.~A.,  {de Bruyn}
  A.~G.,  {Ciardi} B.,  {Jeli{\'c}} V.,  {Koopmans} L.~V.~E.,  {Labropoulos}
  P.,  {Mellema} G.,  {Offringa} A.,  {Pandey} V.~N.,  {Pawlik} A.~H.,
  {Schaye} J.,  {Thomas} R.~M.,    {Yatawatta} S.,  2010, \mnras, pp 612--+

\bibitem[\protect\citeauthoryear{{Harker}, {Zaroubi}, {Thomas}, {Jeli{\'c}},
  {Labropoulos}, {Mellema}, {Iliev}, {Bernardi}, {Brentjens}, {de Bruyn},
  {Ciardi}, {Koopmans}, {Pandey}, {Pawlik}, {Schaye} \& {Yatawatta}}{{Harker}
  et~al.}{2009}]{harker09}
{Harker} G.~J.~A.,  {Zaroubi} S.,  {Thomas} R.~M.,  {Jeli{\'c}} V.,
  {Labropoulos} P.,  {Mellema} G.,  {Iliev} I.~T.,  {Bernardi} G.,  {Brentjens}
  M.~A.,  {de Bruyn} A.~G.,  {Ciardi} B.,  {Koopmans} L.~V.~E.,  {Pandey}
  V.~N.,  {Pawlik} A.~H.,  {Schaye} J.,    {Yatawatta} S.,  2009, \mnras, 393,
  1449

\bibitem[\protect\citeauthoryear{{Ichikawa}, {Barkana}, {Iliev}, {Mellema} \&
  {Shapiro}}{{Ichikawa} et~al.}{2010}]{ichikawa10}
{Ichikawa} K.,  {Barkana} R.,  {Iliev} I.~T.,  {Mellema} G.,    {Shapiro}
  P.~R.,  2010, \mnras, pp 803--+

\bibitem[\protect\citeauthoryear{{Jeli{\'c}}, {Zaroubi}, {Labropoulos},
  {Bernardi}, {de Bruyn} \& {Koopmans}}{{Jeli{\'c}} et~al.}{2010}]{jelic10}
{Jeli{\'c}} V.,  {Zaroubi} S.,  {Labropoulos} P.,  {Bernardi} G.,  {de Bruyn}
  A.~G.,    {Koopmans} L.~V.~E.,  2010, \mnras, 409, 1647

\bibitem[\protect\citeauthoryear{{Jeli{\'c}}, {Zaroubi}, {Labropoulos},
  {Thomas}, {Bernardi}, {Brentjens}, {de Bruyn}, {Ciardi}, {Harker},
  {Koopmans}, {Pandey}, {Schaye} \& {Yatawatta}}{{Jeli{\'c}}
  et~al.}{2008}]{jelic08}
{Jeli{\'c}} V.,  {Zaroubi} S.,  {Labropoulos} P.,  {Thomas} R.~M.,  {Bernardi}
  G.,  {Brentjens} M.~A.,  {de Bruyn} A.~G.,  {Ciardi} B.,  {Harker} G.,
  {Koopmans} L.~V.~E.,  {Pandey} V.~N.,  {Schaye} J.,    {Yatawatta} S.,  2008,
  \mnras, 389, 1319

\bibitem[\protect\citeauthoryear{{Kaiser} \& {Peacock}}{{Kaiser} \&
  {Peacock}}{1991}]{kaiser91}
{Kaiser} N.,  {Peacock} J.~A.,  1991, \apj, 379, 482

\bibitem[\protect\citeauthoryear{{Lacey} \& {Cole}}{{Lacey} \&
  {Cole}}{1993}]{lacey93}
{Lacey} C.,  {Cole} S.,  1993, \mnras, 262, 627

\bibitem[\protect\citeauthoryear{{Lawson}, {Mayer}, {Osborne} \&
  {Parkinson}}{{Lawson} et~al.}{1987}]{lawson87}
{Lawson} K.~D.,  {Mayer} C.~J.,  {Osborne} J.~L.,    {Parkinson} M.~L.,  1987,
  \mnras, 225, 307

\bibitem[\protect\citeauthoryear{{Lidz}, {Oh} \& {Furlanetto}}{{Lidz}
  et~al.}{2006}]{lidz06}
{Lidz} A.,  {Oh} S.~P.,    {Furlanetto} S.~R.,  2006, \apjl, 639, L47

\bibitem[\protect\citeauthoryear{{Lidz}, {Zahn}, {McQuinn}, {Zaldarriaga} \&
  {Hernquist}}{{Lidz} et~al.}{2008}]{lidz08}
{Lidz} A.,  {Zahn} O.,  {McQuinn} M.,  {Zaldarriaga} M.,    {Hernquist} L.,
  2008, \apj, 680, 962

\bibitem[\protect\citeauthoryear{{Liu}, {Tegmark}, {Bowman}, {Hewitt} \&
  {Zaldarriaga}}{{Liu} et~al.}{009b}]{liu09b}
{Liu} A.,  {Tegmark} M.,  {Bowman} J.,  {Hewitt} J.,    {Zaldarriaga} M.,
  {2009b}, \mnras, 398, 401

\bibitem[\protect\citeauthoryear{{Liu}, {Tegmark} \& {Zaldarriaga}}{{Liu}
  et~al.}{2009}]{liu09a}
{Liu} A.,  {Tegmark} M.,    {Zaldarriaga} M.,  2009, \mnras, 394, 1575

\bibitem[\protect\citeauthoryear{{McQuinn}, {Zahn}, {Zaldarriaga}, {Hernquist}
  \& {Furlanetto}}{{McQuinn} et~al.}{2006}]{mcquinn06}
{McQuinn} M.,  {Zahn} O.,  {Zaldarriaga} M.,  {Hernquist} L.,    {Furlanetto}
  S.~R.,  2006, \apj, 653, 815

\bibitem[\protect\citeauthoryear{{Mesinger} \& {Furlanetto}}{{Mesinger} \&
  {Furlanetto}}{2007}]{mesinger07}
{Mesinger} A.,  {Furlanetto} S.,  2007, \apj, 669, 663

\bibitem[\protect\citeauthoryear{{Mesinger}, {Furlanetto} \& {Cen}}{{Mesinger}
  et~al.}{2010}]{mesinger10}
{Mesinger} A.,  {Furlanetto} S.,    {Cen} R.,  2010, ArXiv e-prints

\bibitem[\protect\citeauthoryear{{Morales}, {Bowman} \& {Hewitt}}{{Morales}
  et~al.}{2006a}]{morales05_foregrounds}
{Morales} M.~F.,  {Bowman} J.~D.,    {Hewitt} J.~N.,  2006a, \apj, 648, 767

\bibitem[\protect\citeauthoryear{{Morales}, {Bowman} \& {Hewitt}}{{Morales}
  et~al.}{2006b}]{morales06}
{Morales} M.~F.,  {Bowman} J.~D.,    {Hewitt} J.~N.,  2006b, \apj, 648, 767

\bibitem[\protect\citeauthoryear{{Morales} \& {Hewitt}}{{Morales} \&
  {Hewitt}}{2004}]{morales04}
{Morales} M.~F.,  {Hewitt} J.,  2004, \apj, 615, 7

\bibitem[\protect\citeauthoryear{{Morales} \& {Wyithe}}{{Morales} \&
  {Wyithe}}{2009}]{morales09}
{Morales} M.~F.,  {Wyithe} J.~S.~B.,  2009, ArXiv e-prints

\bibitem[\protect\citeauthoryear{{Oh} \& {Mack}}{{Oh} \& {Mack}}{2003}]{oh03}
{Oh} S.~P.,  {Mack} K.~J.,  2003, \mnras, 346, 871

\bibitem[\protect\citeauthoryear{{Roy}, {Goss} \& {Anantharamaiah}}{{Roy}
  et~al.}{2008}]{roy08}
{Roy} A.~L.,  {Goss} W.~M.,    {Anantharamaiah} K.~R.,  2008, \aap, 483, 79

\bibitem[\protect\citeauthoryear{{Roy}, {Oosterloo}, {Goss} \&
  {Anantharamaiah}}{{Roy} et~al.}{2010}]{roy10}
{Roy} A.~L.,  {Oosterloo} T.,  {Goss} W.~M.,    {Anantharamaiah} K.~R.,  2010,
  \aap, 517, A82+

\bibitem[\protect\citeauthoryear{{Rybicki} \& {Lightman}}{{Rybicki} \&
  {Lightman}}{1979}]{rybicki79}
{Rybicki} G.~B.,  {Lightman} A.~P.,  1979, {Radiative Processes in
  Astrophysics}.
New York: Wiley

\bibitem[\protect\citeauthoryear{{Rybicki} \& {Press}}{{Rybicki} \&
  {Press}}{1992}]{rybicki92}
{Rybicki} G.~B.,  {Press} W.~H.,  1992, \apj, 398, 169

\bibitem[\protect\citeauthoryear{{Santos}, {Cooray} \& {Knox}}{{Santos}
  et~al.}{2005}]{santos05}
{Santos} M.~G.,  {Cooray} A.,    {Knox} L.,  2005, \apj, 625, 575

\bibitem[\protect\citeauthoryear{{Seaquist} \& {Bell}}{{Seaquist} \&
  {Bell}}{1977}]{seaquist77}
{Seaquist} E.~R.,  {Bell} M.~B.,  1977, \aap, 60, L1+

\bibitem[\protect\citeauthoryear{{Shaver}}{{Shaver}}{1978}]{shaver78}
{Shaver} P.~A.,  1978, \aap, 68, 97

\bibitem[\protect\citeauthoryear{{Shaver}, {Churchwell} \& {Rots}}{{Shaver}
  et~al.}{1977}]{shaver77}
{Shaver} P.~A.,  {Churchwell} E.,    {Rots} A.~H.,  1977, \aap, 55, 435

\bibitem[\protect\citeauthoryear{{Shaver}, {Windhorst}, {Madau} \& {de
  Bruyn}}{{Shaver} et~al.}{1999}]{shaver99}
{Shaver} P.~A.,  {Windhorst} R.~A.,  {Madau} P.,    {de Bruyn} A.~G.,  1999,
  \aap, 345, 380

\bibitem[\protect\citeauthoryear{{Shioya}, {Trentham} \& {Taniguchi}}{{Shioya}
  et~al.}{2001}]{shioya01}
{Shioya} Y.,  {Trentham} N.,    {Taniguchi} Y.,  2001, \apjl, 548, L29

\bibitem[\protect\citeauthoryear{{Tegmark}, {Eisenstein}, {Hu} \& {de
  Oliveira-Costa}}{{Tegmark} et~al.}{2000}]{tegmark00}
{Tegmark} M.,  {Eisenstein} D.~J.,  {Hu} W.,    {de Oliveira-Costa} A.,  2000,
  \apj, 530, 133

\bibitem[\protect\citeauthoryear{{Wang}, {Tegmark}, {Santos} \& {Knox}}{{Wang}
  et~al.}{2006}]{wang06}
{Wang} X.,  {Tegmark} M.,  {Santos} M.~G.,    {Knox} L.,  2006, \apj, 650, 529

\bibitem[\protect\citeauthoryear{{Wyithe} \& {Morales}}{{Wyithe} \&
  {Morales}}{2007}]{wyithe07}
{Wyithe} J.~S.~B.,  {Morales} M.~F.,  2007, \mnras, 379, 1647

\bibitem[\protect\citeauthoryear{{Young}, {Claussen}, {Kleinmann}, {Rubin} \&
  {Scoville}}{{Young} et~al.}{1988}]{young88}
{Young} J.~S.,  {Claussen} M.~J.,  {Kleinmann} S.~G.,  {Rubin} V.~C.,
  {Scoville} N.,  1988, \apjl, 331, L81

\bibitem[\protect\citeauthoryear{{Yun}, {Reddy} \& {Condon}}{{Yun}
  et~al.}{2001}]{yun01}
{Yun} M.~S.,  {Reddy} N.~A.,    {Condon} J.~J.,  2001, \apj, 554, 803

\bibitem[\protect\citeauthoryear{{Zaldarriaga}, {Furlanetto} \&
  {Hernquist}}{{Zaldarriaga} et~al.}{2004}]{zald04}
{Zaldarriaga} M.,  {Furlanetto} S.~R.,    {Hernquist} L.,  2004, \apj, 608, 622

\bibitem[\protect\citeauthoryear{{Zhao}, {Anantharamaiah}, {Goss} \&
  {Viallefond}}{{Zhao} et~al.}{1996}]{zhao96}
{Zhao} J.,  {Anantharamaiah} K.~R.,  {Goss} W.~M.,    {Viallefond} F.,  1996,
  \apj, 472, 54

\end{thebibliography}
%\bsp 
\appendix

\section[]{Foreground Fluctuations in Frequency Space}
\label{section:smoothness}

The generic property that all foreground cleaning algorithms use is that foregrounds should be spectrally smooth, allowing the line features of 21cm emission to be picked out. However, given that foregrounds can exceed the signal by $\sim 4-5$ orders of magnitude, it is not immediately obvious that tiny spectral variations in the foreground will not mask the 21cm signal. In this section, we quantify the required level of smoothness, and show that synchrotron and free-free emission from our Galaxy and extragalactic sources indeed satisfy it.

If the foreground is a perfectly smooth power-law $T_{b} \propto \nu^{-\alpha}$, then the 21cm forest can be extracted, since it consists of narrow line features. However, there are inevitable variations in the spectral index with position in the sky as well as frequency, and our ability to subtract the galactic synchrotron foreground is dominated by the uncertainty in the spectral index. This dispersion is of order $\Delta \alpha \sim 0.03-0.1$ at $\sim 200$ MHz \citep{lawson87,banday91}. The required smoothness of the foreground in frequency space is therefore often characterized by uncertainty in the spectral index $\Delta \alpha$ (e.g., \citet{shaver99,dimatteo02,gnedin04}). The temperature uncertainty due to an uncertainty in the spectral index $\Delta \alpha$ is $(\Delta T/T) \approx (\nu/\nu_{o})^{\Delta \alpha} -1 \approx \Delta \alpha (\Delta \nu/\nu)$. The 21cm fluctuations occur over a frequency interval $(\Delta \nu/\nu) \sim r_{\rm ion}/l_{H} \sim 10^{-2}$, where $r_{\rm ion}$ is the characteristic size of ionized/neutral patches and $l_{H}$ is the Hubble length. These correspond to a variation in the spectral index of:
\begin{equation}
\Delta \alpha({\rm reion}) \approx  5 \times 10^{-3} \left( \frac{\Delta T/T_{b}}{5 \times 10^{-5}} \right) \left( \frac{\delta \nu/\nu} {10^{-2}} \right)^{-1}.
\end{equation}
Given the large $\Delta \alpha \sim 0.03-0.1$ variations in the spectral index observed in our Galaxy observed from point to point and over fairly large frequency intervals, $\delta \nu/\nu \sim {\rm few} \times 0.1$, it is not clear that the spectral index can be constrained to the $\sim 10^{-3}$ accuracy required.

How strong will the deviations from power law emission be? Synchrotron emission is a power law because the electron momentum distribution is a power-law (which is generically true, for instance, for Fermi acceleration). An electron momentum distribution $N(\gamma) d\gamma \propto \gamma^{-p} d\gamma$ will produce a synchrotron emission spectrum $F_{\nu} \propto \nu^{-(p-1)/2}$. However, the shorter lifetimes of high-momentum electrons will introduce curvature even for an initially perfect power law. Moreover, turbulence, non-linear scattering, magnetic field strength fluctuations and plasma effects could introduce fluctuations in the momentum distribution. Will this translate into sufficiently large spectral index fluctuations so as to swamp the 21cm signal?

It is easiest to think directly in terms of temperature fluctuations, rather than the spectral index. In order for galactic foregrounds not be a significant source of contamination, we shall simply require that the rms temperature fluctuations smoothed on $\delta \nu\sim 0.1-1$MHz intervals be significantly less than the expected 21cm signal:
\begin{equation}
\langle \Delta T_{fg}^{2} \rangle^{1/2} \ll 10^{-2} {\rm K}.
\end{equation}
The observed emission as a function of frequency $T(\omega)$, is the convolution of the electron momentum distribution $N(\gamma)$ with the emission spectrum of a single electron $E(\omega)$: $T(\omega) \propto N(\gamma) \otimes E(\omega)$. Thus, by the convolution theorem, the power spectrum of temperature fluctuations as a function of smoothing frequency $\Delta \omega_{k}$ is given by:
\begin{equation}
P(k_{\omega})=P(k_{\gamma})W_{k}^{2}
\end{equation}
where $W_{k}$ is the Fourier transform of the emission spectrum $E(\omega)$. The window function $W_{k}$ therefore quantifies the suppression of small-scale temperature fluctuations due to smoothing by the broad emission kernel $E(\omega)$.

\begin{figure}
\includegraphics[width=0.45\textwidth]{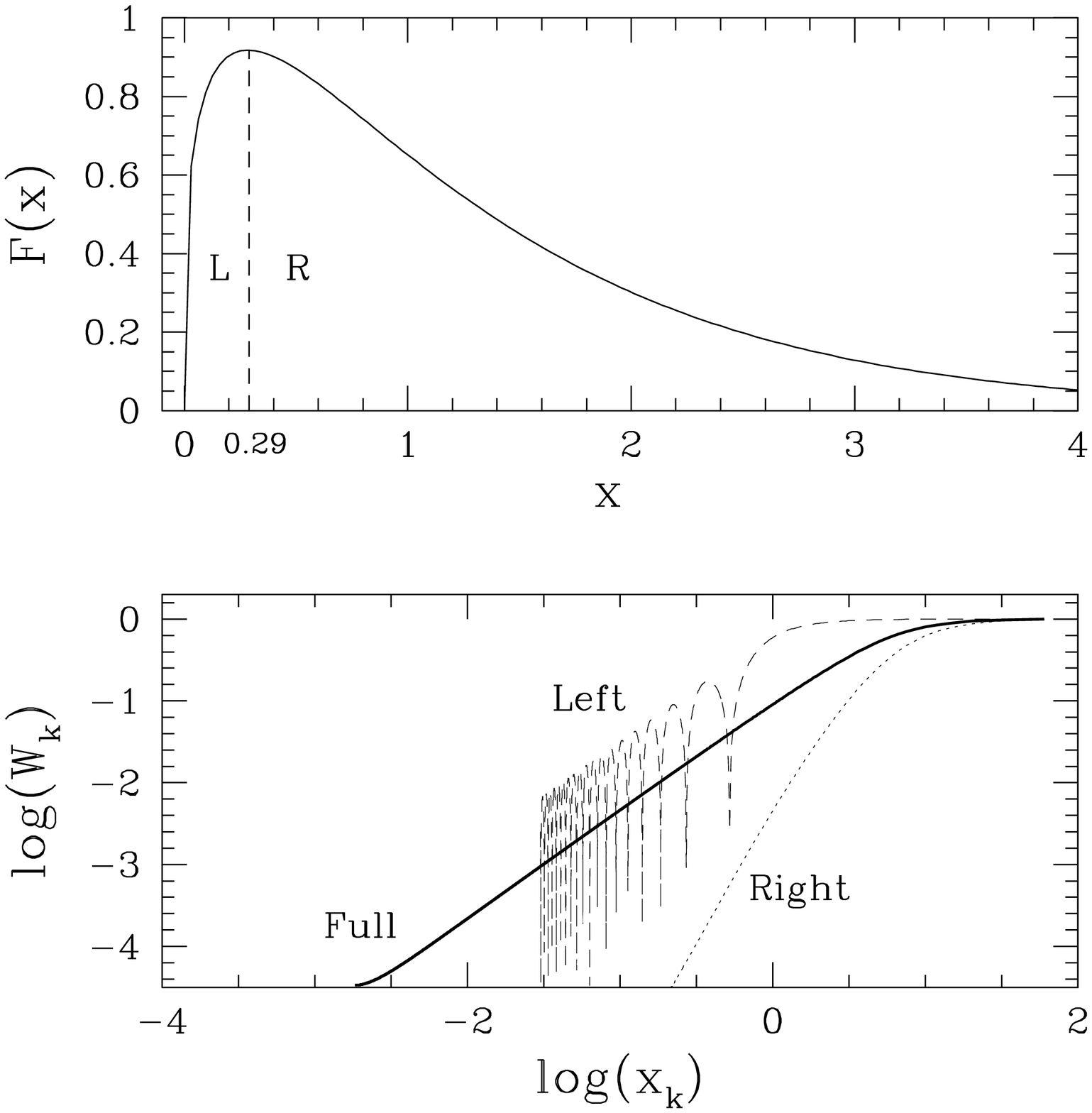}
\caption{{\it Top panel}: Frequency dependence of the synchrotron spectrum of a single electron with Lorentz factor $\gamma$, as given by equation (\ref{eqn:f(x)}), in units of $x\equiv \omega/\omega_{c}$.
The emission peaks at $x_{\rm max}=0.29$; the low frequency portion ('L' for Left) damps more rapidly than the high frequency portion ('R' for Right). {\it Bottom panel}: The window function $W_{k}$, which is the Fourier transform of the emission spectrum $F(x)$, as a function of $x_{k}=1/k$. For the small frequency intervals relevant for the 21cm forest $x_{k} < 10^{-2.5}$, the window function has small amplitude, $W_{k}\sim 10^{-4}$, implying that foreground temperature fluctuations are negligible on these scales. The decay of the window function is dominated by the narrow left portion of the emission spectrum (dashed lines), rather than the broad right portion (dotted lines).}
\label{fig:sync_spectrum}
\end{figure}

Let us now calculate $W_{k}$. The power per unit frequency emitted by each electron is \citep{rybicki79}:
\begin{equation}
E(\omega)=\frac{\sqrt{3} q^{3} B \, {\rm sin} \alpha}{ 2\pi m_{e} c^{2}} F(x)
\end{equation}
where $\alpha$ is the pitch angle between the magnetic field and electron velocity, while
\begin{equation}
F(x)\equiv x \int _{x}^{\infty} K_{5/3}{\xi}d\xi,
\label{eqn:f(x)}
\end{equation}
where $K_{5/3}$ is a modified Bessel function of the second kind, and $x\equiv \omega/\omega_{c}$. The critical frequency $\omega_{c}$ is:
\begin{equation}
\omega_{c}\equiv \frac{3}{2} \gamma^{3} \omega_{B} \, {\rm sin}\alpha = \frac{ 3 \gamma^{2} q B \, {\rm sin}\alpha}{2 m_{e} c}
\end{equation}
where $\gamma$ is the electron Lorentz factor. Note that the critical frequency is larger than the electron gyration frequency $\omega_{B}=qB/\gamma m_{e}c$ by a factor $\sim \gamma^{3}$. The power per frequency interval $d\nu$ is simply $P(\nu)=2\pi P(\omega)$. The function $F(x)$ is shown as the top panel of Figure \ref{fig:sync_spectrum}. Note that it is broad in frequency range: $\Delta \nu \sim \nu$. This implies that fluctuations on scales $\Delta \nu \ll \nu$ will be smoothed out.

The window function $W_{k}$ is simply the absolute value of the Fourier transform of $F(x)$.
%; we normalize it to unity on large scales $k\rightarrow 0$, since on large scales, convolution by a narrow window function will have no effect on the power spectrum.
In the bottom panel of Figure \ref{fig:sync_spectrum}, we plot $W_{k}$ as a function of $x_{k}=1/k \approx \Delta \nu/\nu$. In the bottom panel of Figure \ref{fig:sync_spectrum}, we plot $W_{k}$, as well as the window function for the left and right portion separately. Note that $F(x)$ is asymmetric about its peak at $x_{\rm max}=0.29$ and the left (low-frequency portion) is much narrower than the right (high-frequency) portion. The latter sets the characteristic frequency width of the electron's emission. $W_{k}$ is dominated by the narrow left portion, as seen by the separate window functions for the left and right portion (the "ringing" in Fourier space for the left portion is due to the sharpness of edges). The window function clearly acts as a low-pass filter: all high-frequency fluctuations are strongly damped, since the emission spectrum $E(\omega)$ is so broad. 21cm emission corresponds to $x\sim 10^{-4}/x_{\rm max} \sim 10^{-2.5}$; at these small frequency intervals, $W_{k} \sim 10^{-4}$. Thus, even if there are pathologically large fluctuations in the electron distribution function $\Delta_{\gamma} \equiv \left[\delta N(\gamma)/N(\gamma) \right]_{\rm rms} \sim 0.1$ (averaged over bins of width $\delta \gamma / \gamma \sim \delta \nu / \nu \sim 10^{-4}$), they will be strongly suppressed to yield temperature fluctuations on these scales $\delta T/T \sim W_{k} \Delta_{\gamma} \sim 10^{-5} (\Delta_{\gamma}/0.1)$, smaller than the 21cm signal $\delta T/T \sim 10^{-4}$.  In reality, stochastic fluctuations in the electron distribution function are likely to be {\it much} smaller than $\Delta_{\gamma} \sim 0.1$: Poisson fluctuations are utterly negligible, since one is averaging over a huge number of electrons, so some unusual process (for instance, resonances in momentum space) would be necessary to produce sharp features in the electron momentum distribution. Thus, foreground temperature fluctuations over the narrow frequency intervals associated with 21cm emission are utterly negligible.

We can perform a similar estimate for thermal bremsstrahlung emission,
which is the emission spectrum $E(\nu,v)$ of an electron of velocity
$v$, convolved with a Maxwellian velocity distribution. However, the
emission spectrum of even a mono-energetic electron population is very
broad in frequency space: $E(\nu,v) \propto v^{-1} g_{ff}(v,\nu)$,
where the Gaunt factor $g_{ff}(v,\omega)$ is of order unity from low
frequencies out to $h \nu_{\rm crit} \sim \frac{1}{2} m v^{2}$. The
free-free emission spectrum for a monoenergetic distribution is much
broader than that for synchrotron emission, where $\Delta \nu \sim
\nu_{\rm obs}$: for free-free emission, $\Delta \nu\sim \nu_{\rm crit}
\sim (k_{\rm B}T/h) \sim 100 (T/10^{4} \, {\rm K}) \, {\rm THz}\gg
\nu_{\rm obs} \sim 150$MHz. The window function will decay even more
rapidly at small frequency intervals, and there will be no small-scale
frequency structure for free-free emission, even for a turbulent
plasma.

Since extragalactic sources are dominated by free-free or synchrotron
emission, their sum will also be smooth in frequency space. The only
contaminants which might have sufficient small-scale structure in
frequency space to be confused with the 21cm forest are radio
recombination lines. Preliminary estimates show they are unlikely to
be a significant source of contamination out of the Galactic plane
\citep{shaver99,oh03}, but there could be surprises. Below, we estimate the integrated background of radio recombination lines from star-forming galaxies.

\section[]{The Radio Recombination Line Background}
\label{section:RRL}

In this Appendix, we perform simple estimates of the foreground contamination due to extragalactic radio recombination lines (RRLs), which have frequencies: 
\begin{equation}
\nu = 153 \, \Delta n \left( \frac{n}{350} \right)^{-3} (1+z)^{-1} \, {\rm MHz}. 
\end{equation}
Extragalactic RRLs were first detected from the starbursts in M82 and NGC 253 by \citet{shaver77} and \citet{seaquist77}. It was soon pointed out that stimulated emission from a strong non-thermal background could allow distant radio galaxies and quasars to be seen in RRL emission \cite{shaver78}. This expectation has not been borne out, likely because the volume filling factor of HII regions around radio quasars is small \citep{anantharamaiah93}. Nonetheless, they are detectable in bright nuclear starburst regions in nearby galaxies; to date, there are 15 known extragalactic RRL detections \citep{roy08,roy10}. Models of RRL emission are highly uncertain and sensitive to the unknown gas density and geometry in nuclear regions. At higher frequencies, emission appears to be due to a mixture of spontaneous emission and stimulated emission by free-free continuum within the HII regions, while at lower frequencies stimulated emission by non-thermal continuum can be important. In this paper, we adopt a crude conservative upper bound on RRL emission. We find the level of contamination is so small that more sophisticated estimates are unnecessary. 

We begin by estimating the luminosity due to internal emission (spontaneous and stimulated) in HII regions. Under optically thin conditions, and considering only the strongest $\alpha$ lines ($\Delta n=1$), the flux from an HII region in RRLs is (\citet{shaver78}; also see \citet{gordon_rrl_book}): 
\begin{equation}
\Delta {\rm S_{L}} = \frac{0.72 \nu}{\Delta V T_{e}^{3/2}} \frac{n_{e} M_{\rm HII}}{D^{2}} b_{n} (1-\beta_{n} \tau_{\rm C}/2) \, {\rm mJy}  
\label{eqn:flux_shaver}
\end{equation}
where $n_{e}$(in ${\rm cm^{-3}}$) is the gas density, $M_{\rm HII}$ (in $M_{\odot}$) is the mass of ionized hydrogen, $\Delta V$ (in ${\rm km \, s^{-1}}$) is the line width, $T_{e}$ (in K) is the temperature, $\nu$ (in GHz) is the frequency, and $b_{n} (1-\beta_{n} \tau_{\rm C}/2)$ captures departures from thermal equilibrium\footnote{In particular, $\tau_{\rm C}$ is the continuum (free-free) optical depth, while $b_{\rm n} \equiv N_{\rm n}/N^{*}_{\rm n}$ is the departure coefficient which relates the population of atomic energy level $n$, $N_{n}$, to its LTE value $N_{n}^{*}$. The central line optical depth is then related to the LTE value by $\tau_{\rm L} = b_{\rm n} \beta_{\rm n} \tau_{\rm L}^{*}$, where $\beta_{\rm n} \equiv 1- (kT_{e})/(h \nu)(\Delta b\Delta n)/b_{\rm n}$. Under conditions appropriate for strong stimulated emission, typically $b_{\rm n}\beta_{\rm n} \sim 10-100$.}. The dependence on $n$ is weak as long as $n\gg 1$; see \citet{shaver78} for the full expression. Assuming that the escape fraction of ionizing photons is small, ionization equilibrium in HII regions within the galaxy implies that $\alpha_{\rm B} n^{2} V = \alpha_{\rm B} n (M_{\rm HII}/m_{p}) = \dot{N}_{\rm ion} = 10^{53} \, {\rm s^{-1}} ({\rm SFR}/1 {\rm M_{\odot} \, yr^{-1}})$, which allows us to related the quantity $n_{e} M_{\rm HII}$ in equation (\ref{eqn:flux_shaver}) with the star formation rate (SFR) of a galaxy. Performing this substitution, we obtain the RRL luminosity for internally generated radiation: 
\begin{eqnarray} 
L_{\nu}^{\rm internal} = 1.3 \times 10^{25} \, {\rm erg \, s^{-1} \, Hz^{-1}} \left( \frac{\rm SFR}{1 {\rm M_{\odot} \, yr^{-1}}} \right) \left( \frac{\nu}{1 \, {\rm GHz}} \right) \\ \nonumber
\left(\frac{\Delta V}{200 \, {\rm km \, s^{-1}}} \right)^{-1} \left( \frac{T_{e}}{10^{4} \, {\rm K}} \right)^{-3/2} \left[ \frac{b_{n}(1-\beta_{n} \tau_{C}/2)}{100} \right].  
\end{eqnarray}    
Note that we have adopted an extremely optimistic expression for the boost factor due to stimulated emission, $b_{n}(1-\tau_{\rm C}\beta_{n}/2) \sim 100$. 

Let us proceed to estimate the RRL luminosity due to stimulated emission from external radiation. Consider a collection of compact HII regions around a starburst. One can place a conservative upper bound on stimulated emission by considering an extreme case where $N$ HII regions lie in front of a uniformly distributed background emission $S_{\rm Cbg}$ \citep{zhao96}:
\begin{equation}
\Delta S_{\rm L} = N_{\rm HII}^{\rm los} \left(\frac{\Delta V_{\rm HII}}{\Delta V_{\rm obs}} \right) S_{\rm Cbg} e^{-\tau_{C}} (e^{-\tau_{\rm L}} -1)  
\label{eqn:S_L}
\end{equation}   
where $N_{\rm los}$ is the number of HII regions along the line of sight (generally $N_{\rm los}<1$ observationally, and is required in this formula since we ignore shadowing), $\Delta V_{\rm HII} \sim 20 \, {\rm km \, s^{-1}}$ is the Doppler width of individual HII regions, $\Delta V_{\rm obs}$ is the observed line width (either due to motions of HII regions within the galaxy, or the width of the observing channel), and $\tau_{\rm L},\tau_{\rm C}$ are the line and continuum (free-free) optical depths of each HII region. When $\Delta V_{\rm obs}=\Delta V_{\rm HII}$, and for $\tau_{\rm L},\tau_{\rm C} \ll 1$ (which is usually the case), $\Delta S_{\rm L} \sim -N_{\rm HII}^{\rm los} \tau_{\rm L} S_{\rm Cbg}$, which makes intuitive sense: the line-to-continuum ratio is simply the cumulative optical depth of the line (note that for stimulated emission, $\tau_{\rm L} < 0$). The factor of $\Delta V_{\rm HII}/\Delta V_{\rm obs}$ simply expresses frequency dilution. In this paper, we shall conservatively assume a very high fidicial value of $N_{\rm HII}^{los} \tau_{\rm L} \sim 0.1$. While models of the observed emission from starbursts generally yields $N_{\rm HII}^{\rm los} \tau_{\rm L} \sim 10^{-3}-10^{-4}$ at 8.3 GHz, $\tau_{\rm L} \propto \nu^{-1}$, and hence could be larger at lower frequencies. Note that the ratio of continuum to line optical depths is: 
\begin{equation}
\frac{\tau_{\rm C}}{\tau_{\rm L}} \sim 8 \left( \frac{\Delta V_{\rm HII}}{20 \, {\rm km \, s^{-1}}} \right) \left( \frac {\nu}{150 {\rm MHz}} \right)^{-1.1} \left( \frac{b_{\rm n}\beta_{n}}{100} \right)^{-1} \left( \frac{T_{e}}{10^{4}} \right)^{1.15}.  
\end{equation}
Thus, increasing $\tau_{\rm L}$ to values of order unity will also increase free-free absorption, resulting in unobservable RRL emission. Note also that the single "slab" model of \citet{shaver78,anantharamaiah93,zhao96} simply corresponds to $N_{\rm HII}^{\rm los} \rightarrow 1$, such that $S_{\rm L} \approx \tau_{\rm L} S_{\rm CBg}$. 
 
We now need a model for $S_{\rm Cbg}$. At low frequencies, non-thermal emission dominates over free-free emission, and will be the primary source of stimulated emission. Since the covering fraction of HII regions has been generally found to be small (e.g., in \citet{anantharamaiah93}, $f_{\rm HII} \sim 10^{-6}$), stimulated emission from an AGN is generally unimportant. We model the non-thermal emission to be  primarily synchrotron emission from supernova remnants. The observed correlation between SFR and radio luminosity is \citep{yun01}: 
\begin{equation}
L_{\nu}^{\rm Cbg} = 2.2 \times 10^{28} \, {\rm erg \, s^{-1} \, Hz^{-1}} \left( \frac{\rm SFR}{1 {\rm M_{\odot} \, yr^{-1}}} \right) \left( \frac{\nu}{1 \, {\rm GHz}} \right)^{-0.8}
\end{equation}
Then from equation \ref{eqn:S_L}, we have: 
\begin{eqnarray}
L_{\nu}^{\rm external} = 2.2 \times 10^{26} \, {\rm erg \, s^{-1} \, Hz^{-1}} \, \left( \frac{\nu}{1 \, {\rm GHz}} \right)^{-0.8} \left(\frac{\tau_{\rm L} N_{\rm HII}^{\rm los}}{0.1} \right)  \\ \nonumber
\left( \frac{\rm SFR}{1 {\rm M_{\odot} \, yr^{-1}}} \right) \left(\frac{\Delta V_{\rm HII}}{20 \, {\rm km \, s^{-1}}} \right) \left(\frac{\Delta V_{\rm obs}}{200 \, {\rm km \, s^{-1}}} \right)^{-1} \left( \frac{1+z}{3} \right)^{-0.8}.
\end{eqnarray}
We stress that these fiducial parameters ($b_{n}(1-\tau_{\rm C}\beta_{n}/2) \sim 100$, $\tau_{\rm L} N_{\rm HII}^{\rm los} \sim 0.1$) are by design significant overestimates to conservatively place an upper bound on RRL emission. For instance, from published values of SFR$\sim 270 \, {\rm M_{\odot} \, yr^{-1}}$ \citep{shioya01}, $\sim 4 {\, \rm M_{\odot} \, yr^{-1}}$ \citep{boissier05}, $\sim 6 \, {\rm M_{\odot} \, yr^{-1}}$ \citep{young88} and measured line widths of $320, 95, 200 \, {\rm km \, s^{-1}}$ \citep{zhao96} for Arp 220, M83, and NGC 2146 respectively, our model predicts line flux densities of ${\rm S_{L}} \sim 4.0, 44, 4.7$ mJy for the H92$\alpha$ line at 8.3 GHz, compared to observed values of ${\rm S_{L}} \sim 0.4, 0.8, 0.36$mJy \citep{zhao96}. For the H165$\alpha$ and H167$\alpha$ lines at 1.4 GHz, we predict a flux of $\sim 5$mJy, while an upper bound on the peak line density is $< 0.25$ mJy ($3\sigma$; \citet{anantharamaiah00}). Thus, as is our intent, we consistently overestimate the RRL flux by at least an order of magnitude (we do {\it not} advocate using these parameters to estimate the detectability of RRLs!). 

It is clear that since $L_{\nu}^{\rm int} \propto \nu$, while $L_{\nu}^{\rm ext} \propto \nu^{-0.8}$, they dominate at high and low frequencies respectively. In our model, the cross-over point is at a rest frame frequency of $\nu \sim 4.8$GHz. Since we are concerned with RRLs at rest frame frequencies well below 1.4 GHz, the RRL emission stimulated by external non-thermal sources strongly dominates. In terms of parameters appropriate for 21cm experiments, the RRL luminosity at low frequencies is then: 
\begin{eqnarray}
L_{\nu}^{\rm external} = 4.2 \times 10^{25} \, {\rm erg \, s^{-1} \, Hz^{-1}} \, \left( \frac{\nu_{\rm obs}}{150 \, {\rm MHz}} \right)^{0.2} \left(\frac{\tau_{\rm L} N_{\rm HII}^{\rm los}}{0.1} \right)  \\ \nonumber
\left( \frac{\rm SFR}{1 {\rm M_{\odot} \, yr^{-1}}} \right) \left(\frac{\Delta V_{\rm HII}}{20 \, {\rm km \, s^{-1}}} \right) \left(\frac{\Delta \nu}{1 \, {\rm MHz}} \right)^{-1} \left( \frac{1+z}{3} \right)^{-0.8}.
\end{eqnarray}
where we have substituted $\Delta \nu$ for $\Delta V_{\rm obs}$; in general the line will always be unresolved by the detector unless bandwidths are extremely small, $\Delta \nu \sim 0.1$ MHz. 

We can now estimate the typical brightness temperature perturbation due to an integrated background of RRL emitters. Assuming a comoving star formation rate of $\epsilon_{\rm SFR}$, and the Rayleigh Jeans approximation $T_{\rm L} = c^{2} S_{\rm L}/(2 \, k_{\rm B} \nu^{2} \theta^{2})$, where $\theta$ is the beamsize, we obtain: 
\begin{eqnarray}
\bar{T}_{\rm L} = 1.8 \times 10^{-5} \, {\rm K} \, \left( \frac{\epsilon_{\rm SFR}}{0.1 {\rm M_{\odot} \, yr^{-1} \, Mpc^{-3}}} \right) \left(\frac{\tau_{\rm L} N_{\rm HII}^{\rm los}}{0.1} \right)  \\ \nonumber
\left( \frac{\nu_{\rm obs}}{150 \, {\rm MHz}} \right)^{-2.8} 
 \left(\frac{\Delta V_{\rm HII}}{20 \, {\rm km \, s^{-1}}} \right) \left(\frac{\Delta \nu}{1 \, {\rm MHz}} \right)^{-1} \left( \frac{1+z}{3} \right)^{-3.3}.
\end{eqnarray}
where $z$ refers to the redshift of the RRL emitting galaxies, and not of 21cm emission. Note that this is independent of the bandwidth $\Delta \nu$ or the beamsize $\theta$, since $T_{\rm L}$ has units of surface brightness per frequency interval (increasing $\Delta \nu, \theta$ will increase the RRL luminosity of a region, but it will be spread out over larger angular and frequency intervals). However, the large comoving volume $V_{\rm com} \sim 12^{3} (\Delta \nu/\nu_{\rm obs}/(1/150)) (\theta/5^{\prime})^{2}$Mpc$^{3}$ in a 21cm experiment's field of view is important in estimating rms temperature fluctuations $\delta T_{\rm L}$. Since rms density fluctuations are well below unity on these scales ($\sigma_{8} \sim 0.8$ at $z=0$), an enormous bias factor $b > 10^{3}$ is required for fluctuations in the comoving star formation rate $\delta \epsilon_{\rm SFR}$ to produce an observationally relevant cosmological RRL signal, $\delta T_{\rm L} \sim 10^{-2}$K. Such large bias is obviously unrealistic.  

Thus, it appears fairly robustly that the RRL background is unlikely to be important. Physically, we can understand why this is the case. We modelled $L_{\nu}^{\rm RRL} \sim {\tau_{\rm L} N_{\rm HII}^{\rm los}} L_{\nu}^{\rm Cbg}$ where ${\tau_{\rm L} N_{\rm HII}^{\rm los}} \sim 0.1$, so naively one might think that the RRL foreground is $\sim 10\%$ of the radio continuum foreground (which we know significantly exceeds the 21cm signal). However, whereas all continuum sources along a line of sight contribute, for a specific recombination line, RRL emitters from only a tiny redshift slice contribute\footnote{It is possible that RRLs at different frequencies and thus redshifts can contribute to a fixed bandpass, but they will add incoherently and the signal will be small. The RRL foreground is really only a danger if emission from a given line will be large, since only a small fraction of starbursts exhibit stimulated emission.}. Thus, the very reason why RRLs might pose a danger---the fact that they are localized in frequency space---is also the reason for their small amplitude, since that implies localization in redshift space. Our only note of caution is that observations of extragalactic RRLs have taken place at significantly higher frequencies (e.g., at 1.4, 8.1, 84, 96 and 207 GHz; \citet{anantharamaiah00}); emission at significantly lower frequencies may have different physics. In particular, since $\tau_{\rm L} \propto \nu^{-1}$, lower frequency RRL emission is sensitive to lower density HII regions. However, it is difficult to see how one can vitiate the conservative upper bound we have placed. The matter could be quickly put to rest by measuring the RRL intensities of radio galaxies and starbursts at these frequencies.  

%%\label{lastpage}

\end{document}